\pdfoutput=1
\documentclass[letterpaper,english,11pt,aps,letter,superscriptaddress,nofootinbib,floatfix,pre]{revtex4}
\usepackage[T1]{fontenc}
\usepackage[latin1]{inputenc}
\usepackage{soul}
\usepackage{amsmath}
\usepackage{xcolor}
\usepackage{pdfcolmk}
\usepackage{graphicx}
\usepackage{amssymb}

\makeatletter

\providecolor{lyxadded}{rgb}{1,0,0}
\providecolor{lyxdeleted}{rgb}{0,0,1}

\usepackage{float}
\floatstyle{ruled}
\newfloat{algorithm}{tbp}{loa}
\floatname{algorithm}{Algorithm}
\renewcommand{\citet}[1]{\cite{#1}}

\usepackage{ae,aecompl}
\usepackage{subfigure}

\newenvironment{Algorithm}[1] {
  \refstepcounter{algorithm} 
  \vspace{1ex} \hrule \vspace{1ex} 
  \center{\textbf{Algorithm \thealgorithm : }#1}
  \vspace{1ex} \hrule \vspace{1ex}
  \small
} 
{ \normalsize \vspace{1ex} \hrule \vspace{1ex} }

\makeatother

\usepackage{babel}

\begin{document}

\title{A hybrid particle-continuum method for hydrodynamics of complex fluids}

\author{Aleksandar Donev}

\email{aleks.donev@gmail.com}

\affiliation{Center for Computational Science and Engineering, Lawrence Berkeley
National Laboratory, Berkeley, CA, 94720}

\author{John B. Bell}

\affiliation{Center for Computational Science and Engineering, Lawrence Berkeley
National Laboratory, Berkeley, CA, 94720}

\author{Alejandro L. Garcia}

\affiliation{Department of Physics and Astronomy, San Jose State University, San
Jose, California, 95192}

\author{Berni J. Alder}

\affiliation{Lawrence Livermore National Laboratory, P.O.Box 808, Livermore, CA
94551-9900}

\begin{abstract}
A previously-developed hybrid particle-continuum method {[}\emph{J.
B. Bell, A. Garcia and S. A. Williams, SIAM Multiscale Modeling and
Simulation, 6:1256-1280, 2008}] is generalized to dense fluids and
two and three dimensional flows. The scheme couples an explicit fluctuating
compressible Navier-Stokes solver with the Isotropic Direct Simulation
Monte Carlo (DSMC) particle method {[}\emph{A. Donev and A. L. Garcia
and B. J. Alder, ArXiv preprint 0908.0510}]. To achieve bidirectional
dynamic coupling between the particle (microscale) and continuum (macroscale)
regions, the continuum solver provides state-based boundary conditions
to the particle subdomain, while the particle solver provides flux-based
boundary conditions for the continuum subdomain. This type of coupling
ensures both state and flux continuity across the particle-continuum
interface analogous to coupling approaches for deterministic parabolic
partial differential equations; here, when fluctuations are included,
a small (< 1\%) mismatch is expected and observed in the mean density
and temperature across the interface. By calculating the dynamic structure
factor for both a {}``bulk'' (periodic) and a finite system, it
is verified that the hybrid algorithm accurately captures the propagation
of spontaneous thermal fluctuations across the particle-continuum
interface. The equilibrium diffusive (Brownian) motion of a large
spherical bead suspended in a particle fluid is examined, demonstrating
that the hybrid method correctly reproduces the velocity autocorrelation
function of the bead but only if thermal fluctuations are included
in the continuum solver. Finally, the hybrid is applied to the well-known
adiabatic piston problem and it is found that the hybrid correctly
reproduces the slow non-equilibrium relaxation of the piston toward
thermodynamic equilibrium but, again, only the continuum solver includes
stochastic (white-noise) flux terms. These examples clearly demonstrate
the need to include fluctuations in continuum solvers employed in
hybrid multiscale methods.
\end{abstract}
\maketitle
\newcommand{\Cross}[1]{\left|\boldsymbol{#1}\right|_{\times}}
\newcommand{\CrossL}[1]{\left|\boldsymbol{#1}\right|_{\times}^{L}}
\newcommand{\CrossR}[1]{\left|\boldsymbol{#1}\right|_{\times}^{R}}
\newcommand{\CrossS}[1]{\left|\boldsymbol{#1}\right|_{\boxtimes}}

\newcommand{\V}[1]{\boldsymbol{#1}}
\newcommand{\M}[1]{\boldsymbol{#1}}
\newcommand{\D}[1]{\Delta#1}
\newcommand{\grad}{\boldsymbol{\nabla}}
\newcommand{\Set}[1]{\mathbb{#1}}

\newcommand{\eij}{\left\{  i,j\right\}  }
\newcommand{\Wi}{\mbox{Wi}}

\newcommand{\modified}[1]{\textcolor{red}{#1}}
\newcommand{\deleted}[1]{\textcolor{red}{#1}}
\newcommand{\added}[1]{\textcolor{red}{#1}}

\section{Introduction}

With the increased interest in nano- and micro-fluidics, it has become
necessary to develop tools for hydrodynamic calculations at the atomistic
scale \citet{ParticleMesoscaleHydrodynamics,TripleScale_Rafael,MultiscaleMicrofluidics_Review}.
While the Navier-Stokes-Fourier continuum equations have been surprisingly
successful in modeling microscopic flows \citet{Microfluidics_Review},
there are several issues present in microscopic flows that are difficult
to account for in models relying on a purely PDE approximation. For
example, it is well known that the Navier-Stokes equations fail to
describe flows in the kinetic regions (large Knudsen number flows)
that appear in small-scale gas flows \citet{GasFlows_Nicolas}. It
is also not \emph{a priori} obvious how to account for the bidirectional
coupling between the flow and embedded micro-geometry or complex boundaries.
Furthermore, it is not trivial to include thermal fluctuations in
Navier-Stokes solvers \citet{FluctuatingHydro_Garcia,FluctuatingHydro_Coveney,StagerredFluctHydro,LLNS_S_k},
and in fact, most of the time the fluctuations are omitted even though
they can be important at instabilities \citet{FluidMixing_DSMC} or
in driving polymer dynamics \citet{LatticeBoltzmann_Polymers,StochasticImmersedBoundary}.
An alternative is to use particle-based methods, which are explicit
and unconditionally stable, robust, and simple to implement. The fluid
particles can be directly coupled to the microgeometry, for example,
they can directly interact with the beads of a polymer chain. Fluctuations
are naturally present and can be tuned to have the correct spatio-temporal
correlations.

Several particle methods have been described in the literature, such
as molecular dynamics (MD) \citet{PolymerShear_MD}, Direct Simulation
Monte Carlo (DSMC) \citet{DSMCReview_Garcia}, dissipative particle
dynamics (DPD) \citet{DPD_DNA}, and multi-particle collision dynamics
(MPCD) \citet{DSMC_MPCD_Gompper,DSMC_MPCD_MD_Kapral}. Here we use
the Isotropic DSMC (I-DSMC) stochastic particle method described in
Ref. \citet{SHSD}. In the I-DSMC method, deterministic interactions
between the particles are replaced with stochastic momentum exchange
(collisions) between nearby particles. The I-DSMC method preserves
the essential hydrodynamic properties of expensive MD: local momentum
conservation, linear momentum exchange on length scales comparable
to the particle size, and a similar fluctuation spectrum. At the same
time, the I-DSMC fluid is ideal and structureless, and as such is
much simpler to couple to a continuum solver.

\begin{figure}[tbph]
\begin{centering}
\includegraphics[width=0.5\textwidth]{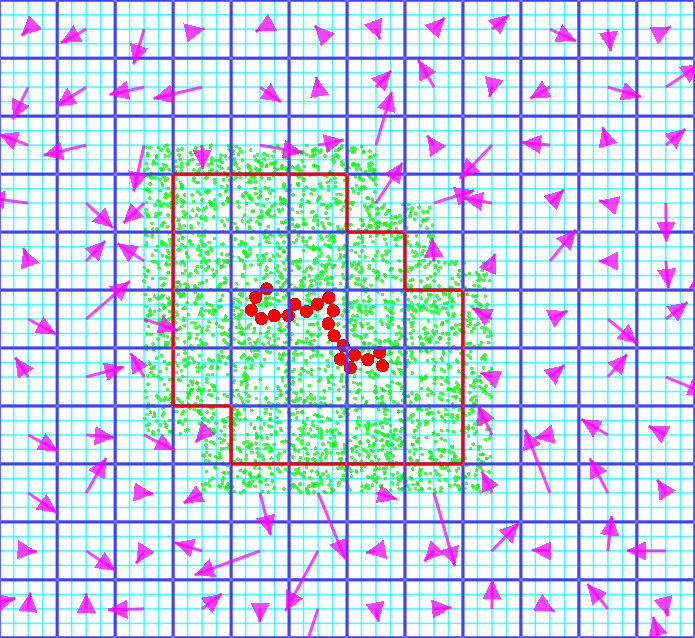}
\par\end{centering}

\caption{\label{HybridIllustration}A two-dimensional illustration of the use
of the MAR hybrid to study a polymer chain (larger red circles) suspended
in an I-DSMC particle fluid. The region around the chain is filled
with particles (smaller green circles), while the remainder is handled
using a fluctuating hydrodynamic solver. The continuum (macro) solver
grid is shown (thicker blue lines), along with the (micro) grid used
by the particle method (thinner blue lines). The fluctuating velocities
in the continuum region are shown as vectors originating from the
center of the corresponding cell (purple). The interface between the
particle and continuum regions is highlighted (thicker red line).}

\end{figure}

However, even particle methods with coarse-grained dynamics, such
as I-DSMC, lack the efficiency necessary to study realistic problems
because of the very large numbers of particles needed to fill the
required computational domain. Most of the computational effort in
such a particle method would, however, be expended on particles far
away from the region of interest, where a description based on the
Navier-Stokes equations is likely to be adequate. Hybrid methods are
a natural candidate to combine the best features of the particle and
continuum descriptions. A particle method can be used in regions where
the continuum description fails or is difficult to implement, such
as near suspended structures or complex boundaries, and a more efficient
continuum description can be used to around the particle domain, as
illustrated in Fig. \ref{HybridIllustration}. This type of hybrid
algorithm fits in the Multi-Algorithm Refinement (MAR) simulation
approach to the modeling and simulation of multiscale problems \citet{AMAR_Burgers,AMAR_DSMC,FluctuatingHydro_AMAR}.
MAR combines two or more algorithms, each of which is appropriate
for a different scale regime. The general idea is to perform detailed
calculations using an accurate but expensive (e.g., particle) algorithm
in a small region, and couple this computation to a simpler, less
expensive (e.g., continuum) method applied to the rest. The challenge
is to ensure that the numerical coupling between the different levels
of the algorithm hierarchy, and especially the coupling of the particle
and continuum computations, is self-consistent, stable, and most importantly,
does not adversely impact the underlying physics.

A crucial feature of our hybrid algorithm is that the continuum solver
includes thermal fluctuations in the hydrodynamic equations consistent
with the particle dynamics, as previously investigated in one dimension
in Ref. \citet{FluctuatingHydro_AMAR}. Thermal fluctuations play
an important role in describing the state of the fluid at microscopic
and mesoscopic scales, especially when investigating systems where
the microscopic fluctuations drive a macroscopic phenomenon such as
the evolution of instabilities, or where the thermal fluctuations
drive the motion of suspended microscopic objects in complex fluids.
Some examples in which spontaneous fluctuations can significantly
affect the dynamics include the breakup of droplets in jets \citet{BreakupNanojets,Nanojets_Eggers},
Brownian molecular motors \citet{BrownianMotors}, Rayleigh-Bernard
convection \citet{RayleighBernard_Fluctuations}, Kolmogorov flows
\citet{Kolmogorov_Fluctuations_1,Kolmogorov_Fluctuations_2}, Rayleigh-Taylor
mixing \citet{FluidMixing_DSMC}, and combustion and explosive detonation
\citet{Detonation_Fluctuations}. In our algorithm, the continuum
solver is a recently-developed three-stage Runge-Kutta integration
scheme for the Landau-Lifshitz Navier-Stokes (LLNS) equations of fluctuating
hydrodynamics in three dimensions \citet{LLNS_S_k}, although other
finite-volume explicit schemes can trivially be substituted.

As summarized in Section \ref{SectionCouplingOverview}, the proposed
hybrid algorithm is based on a fully dynamic bidirectional state-flux
coupling between the particle and continuum regions. In this coupling
scheme the continuum method provides state-based boundary conditions
to the particle subdomain through reservoir particles inserted at
the boundary of the particle region at every particle time step. During
each continuum time step a certain number of particle time steps are
taken and the total particle flux through the particle-continuum interface
is recorded. This flux is then imposed as a flux-based boundary condition
on the continuum solver, ensuring strict conservation \citet{AMAR_DSMC,FluctuatingHydro_AMAR}.
Section \ref{SectionCoupling} describes the technical details of
the hybrid algorithm, focusing on components that are distinct from
those described in Refs. \citet{AMAR_DSMC,FluctuatingHydro_AMAR}.
Notably, the use of the Isotropic DSMC particle method instead of
the traditional DSMC method requires accounting for the interactions
among particles that are in different continuum cells.

In Section \ref{sec:Results} we thoroughly test the hybrid scheme
in both equilibrium and non-equilibrium situations, and in both two
and three dimensions. In Section \ref{SectionMismatch} we study the
continuity of density and temperature across the particle-continuum
interface and identify a small mismatch of order $1/N_{0}$, where
$N_{0}$ is the number of particles per continuum cell, that can be
attributed to the use of fluctuating values instead of means. In Section
\ref{Section_S_kw} we compute dynamic structure factors in periodic
({}``bulk'') and finite quasi two- and one-dimensional systems and
find that the hybrid method seamlessly propagates thermal fluctuations
across the particle-continuum interface.

In Section \ref{SectionVACF} we study the diffusive motion of a large
spherical buoyant bead suspended in a bead of I-DSMC particles in
three dimensions by placing a mobile particle region around the suspended
bead. This example is of fundamental importance in complex fluids
and micro-fluidics, where the motion of suspended objects such as
colloidal particles or polymer chains has to be simulated. Fluctuations
play a critical role since they are responsible for the diffusive
motion of the bead. The velocity-autocorrelation function (VACF) of
a diffusing bead has a well-known power law tail of hydrodynamic origin
and its integral determines the diffusion coefficient. Therefore,
computing the VACF is an excellent test for the ability of the hybrid
method to capture the influence of hydrodynamics on the macroscopic
properties of complex fluids.

Finally, in Section \ref{SectionPiston} we study the slow relaxation
toward thermal equilibrium of an adiabatic piston with the particle
region localized around the piston. In the formulation that we consider,
the system is bounded by adiabatic walls on each end and is divided
into two chambers by a mobile and thermally insulating piston that
can move without friction. We focus on the case when the initial state
of the system is in mechanical equilibrium but not thermodynamic equilibrium:
the pressures on the two sides are equal but the temperatures are
not. Here we study the slow equilibration of the piston towards the
state of thermodynamic equilibrium, which happens because asymmetric
fluctuations on the two sides of the piston slowly transfer energy
from the hotter to the colder chamber.

We access the performance of the hybrid by comparing to purely particle
simulations, which are assumed to be {}``correct''. Unlike in particle
methods, in continuum methods we can trivially turn off fluctuations
by not including stochastic fluxes in the Navier-Stokes equations.
By turning off fluctuations in the continuum region we obtain a \emph{deterministic
hybrid} scheme, to be contrasted with the \emph{stochastic hybrid}
scheme in which fluctuations in the continuum region are consistent
with those in the particle region. By comparing results between the
deterministic and stochastic hybrid we are able to assess the importance
of fluctuations. We find that the deterministic hybrid gives the wrong
long-time behavior for both the diffusing spherical bead and the adiabatic
piston, while the stochastic hybrid correctly reproduces the purely
particle runs at a fraction of the computational cost. These examples
demonstrate the need to include thermal fluctuations in the continuum
solvers in hybrid particle-continuum methods.

\section{\label{SectionCouplingOverview}Brief Overview of the Hybrid Method}

In this section we briefly introduce the basic concepts behind the
hybrid method, delegating further technical details to later sections.
Our scheme is based on an Adaptive Mesh and Algorithm Refinement (AMAR)
methodology developed over the last decade in a series of works in
which a DSMC particle fluid was first coupled to a deterministic compressible
Navier-Stokes solver in three dimensions \citet{AMAR_DSMC,AMAR_DSMC_SAMRAI},
and then to a stochastic (fluctuating) continuum solver in one dimension
\citet{FluctuatingHydro_AMAR}. This section presents the additional
modifications to the previous algorithms necessary to replace the
traditional DSMC particle method with the isotropic DSMC method \citet{SHSD},
and a full three-dimensional dynamic coupling of a complex particle
fluid \citet{DSMC_AED} with a robust fluctuating hydrodynamic solver
\citet{LLNS_S_k}. These novel techniques are discussed further in
Section \ref{SectionCoupling}.

Next, we briefly describe the two components of the hybrid, namely,
the particle \emph{microscopic} model and the continuum \emph{macroscopic}
solver, and then outline the domain decomposition used to couple the
two, including a comparison with other proposed schemes. Both the
particle algorithm and the macroscopic solver have already been described
in detail in the literature, and furthermore, both can easily be replaced
by other methods. Specifically, any variant of DSMC and MPCD can be
used as a particle algorithm, and any explicit finite-volume method
can be used as a continuum solver. For this reason, in this paper
we focus on the coupling algorithm.

\subsection{Particle Model}

The particle method that we employ is the Isotropic DSMC (I-DSMC)
method, a \emph{dense fluid}%
\footnote{Note that by a dense fluid we mean a fluid where the mean free path
is small compared to the typical fluid inter-atomic distance.%
} generalization of the Direct Simulation Monte Carlo (DSMC) algorithm
for rarefied gas flows \citet{DSMCReview_Garcia} . The I-DSMC method
is described in detail in Ref. \citet{SHSD} and here we only briefly
summarize it. It is important to note that, like the traditional DSMC
fluid, the I-DSMC fluid is an \emph{ideal} fluid, that is, it has
the equation of state (EOS) and structure of an ideal gas; it can
be thought of as a viscous ideal gas. As we will see shortly, the
lack of structure in the I-DSMC fluid significantly simplifies coupling
to a continuum solver while retaining many of the salient features
of a dense fluid.

In the I-DSMC method, the effect of interatomic forces is replaced
by stochastic collisions between nearby particles. The interaction
range is controlled via the \emph{collision diameter} $D$ or, equivalently,
the density (hard-sphere volume fraction) $\phi=\pi ND^{3}/(6V)$,
where $N$ is the total number of particles in the simulation volume
$V$. The strength of the interaction is controlled through the dimensionless
\emph{cross-section prefactor} $\chi$, which is on the order of unity
\citet{SHSD}. Stochastic collisions are processed at the end of every
particle time step of duration $\D{t}_{P}$, and in-between collisions
each particle $i$ is streamed advectively with constant velocity
$\V{\upsilon}_{i}=\dot{\V{r}}_{i}$. The spatial domain of the simulation,
typically a rectangular region, is divided into \emph{micro cells}
of length $L_{c}\gtrapprox D$, which are used to efficiently identify
all particles that are within distance $L_{c}$ of a given particle
by searching among the particles in neighboring cells (each cell has
$3^{d}$ neighboring cells, counting itself, where $d$ is the spatial
dimension). At each time step, \emph{conservative collisions} occur
between randomly-chosen pairs of particles that are closer than a
distance $D$ apart; specifically, a random amount of momentum and
energy is exchanged between the two particles with the probabilities
of the collision outcomes obeying detailed balance. The I-DSMC algorithm
can be viewed as a stochastic alternative to deterministic hard-sphere
molecular dynamics (MD) \citet{EventDriven_Alder}, where hard spheres
of diameter $D$ collide when they touch.

In addition to the fluid (I-DSMC) particles there may be a number
of additional spherical particles, which we refer to as \emph{beads},
suspended in the I-DSMC fluid. The beads interact with each other
and the fluid particles either deterministically, as \emph{hard} spheres
impermeable to and colliding with touching particles, or stochastically,
as \emph{permeable} spheres colliding stochastically with overlapping
particles. Note that the sphere radii used for determining the fluid-fluid,
fluid-bead, and bead-bead interaction distances may, in general, be
different. Combining deterministic hard-sphere collisions with stochastic
collisions requires a mixed time-driven and event-driven treatment,
as in the Stochastic Event-Driven MD (SEDMD) algorithm developed in
Ref. \citet{DSMC_AED}. We will use the SEDMD algorithm in the examples
presented in this paper.

We have also developed an alternative purely time-driven fluid-bead
coupling in which the fluid is allowed to permeate the beads and all
of the particle interactions are stochastic. This leads to a much
simpler algorithm that can also easily be parallelized. This distinction
between hard and permeable spheres has analogues in other methods
for complex fluids. For example, in Lattice-Boltzmann simulations
\citet{LB_SoftMatter_Review}, beads can either be modeled as hard
spheres (using a bounce-back collision rule at the lattice sites on
the surface of the bead), or, more efficiently and commonly, as permeable
spheres that let the fluid pass through them but experience a frictional
force due to the fluid motion (exerting the opposite force back on
the lattice sites they overlap with).

\subsection{Continuum Model}

At length scales and time scales larger than the molecular ones, the
dynamics of the particle fluid can be coarse grained \citet{LLNS_Mori,LLNS_Espanol}
to obtain evolution equations for the slow macroscopic variables.
Specifically, we consider the continuum conserved fields \begin{equation}
\V{U}(\V{r},t)=\left[\begin{array}{c}
\rho\\
\V{j}\\
e\end{array}\right]\cong\widetilde{\V{U}}(\V{r},t)=\sum_{i}\left[\begin{array}{c}
m_{i}\\
\V{p}_{i}\\
e_{i}\end{array}\right]\delta\left[\V{r}-\V{r}_{i}(t)\right]=\sum_{i}\left[\begin{array}{c}
1\\
\V{\upsilon}_{i}\\
\upsilon_{i}^{2}/2\end{array}\right]m_{i}\delta\left[\V{r}-\V{r}_{i}(t)\right],\label{U_r_t_def}\end{equation}
where the \emph{conserved variables}, namely the densities of mass
$\rho$, momentum $\V{j}=\rho\V{v}$, and energy $e=\epsilon(\rho,T)+\frac{1}{2}\rho v^{2}$,
can be expressed in terms of the \emph{primitive variables}, mass
density $\rho$, velocity $\V{v}$ and temperature $T$; here $\epsilon$
is the internal energy density. Here the symbol $\cong$ means that
we consider a stochastic field $\V{U}(\V{r},t)$ that approximates
the behavior of the true atomistic configuration $\widetilde{\V{U}}(\V{r},t)$
over long length and time scales (compared to atomistic scales) in
a certain integral average sense; notably, for sufficiently large
cells the integral of $\V{U}(\V{r},t)$ over the cell corresponds
to the total particle mass, momentum and kinetic energy contained
inside the cell.

The evolution of the field $\V{U}(\V{r},t)$ is modeled with the Landau-Lifshitz
Navier-Stokes (LLNS) system of stochastic partial differential equations
(SPDEs) in $d$ dimensions, given in conservative form by\begin{equation}
\partial_{t}\V{U}=-\grad\cdot\left[\V{F}(\V{U})-\V{\mathcal{Z}}(\V{U},\V{r},t)\right],\label{LLNS_general}\end{equation}
where the deterministic flux is taken from the traditional compressible
Navier-Stokes-Fourier equations, \[
\V{F}(\V{U})=\left[\begin{array}{c}
\rho\V{v}\\
\rho\V{v}\V{v}^{T}+P\M{I}-\M{\sigma}\\
(e+P)\V{v}-\left(\M{\sigma}\cdot\V{v}+\M{\xi}\right)\end{array}\right],\]
where $P=P(\rho,T)$ is the pressure, the viscous stress tensor is
$\M{\sigma}=2\eta\left[\frac{1}{2}(\grad\V{v}+\grad\V{v}^{T})-\frac{\left(\grad\cdot\V{v}\right)}{d}\M{I}\right]$
(we have assumed zero bulk viscosity), and the heat flux is $\M{\xi}=\mu\grad T$.
As postulated by Landau-Lifshitz \citet{Landau:Fluid,LLNS_Espanol},
the \emph{stochastic flux} \[
\V{\mathcal{Z}}=\left[\begin{array}{c}
\V{0}\\
\M{\Sigma}\\
\M{\Sigma}\cdot\V{v}+\M{\Xi}\end{array}\right]\]
is composed of the stochastic stress tensor $\M{\Sigma}$ and stochastic
heat flux vector $\M{\Xi}$, assumed to be mutually uncorrelated random
Gaussian fields with a covariance\begin{align}
\left\langle \M{\Sigma}(\V{r},t)\M{\Sigma}^{\star}(\V{r}^{\prime},t^{\prime})\right\rangle = & \M{C}_{\M{\Sigma}}\delta(t-t^{\prime})\delta(\V{r}-\V{r}^{\prime})\mbox{, where }C_{ij,kl}^{(\M{\Sigma})}=2\bar{\eta}k_{B}\overline{T}\left(\delta_{ik}\delta_{jl}+\delta_{il}\delta_{jk}-\frac{2}{d}\delta_{ij}\delta_{kl}\right)\nonumber \\
\left\langle \M{\Xi}(\V{r},t)\M{\Xi}^{\star}(\V{r}^{\prime},t^{\prime})\right\rangle = & \M{C}_{\M{\Xi}}\delta(t-t^{\prime})\delta(\V{r}-\V{r}^{\prime})\mbox{, where }C_{i,j}^{(\M{\Xi})}=2\bar{\mu}k_{B}\overline{T}\delta_{ij},\label{stoch_flux_covariance}\end{align}
where overbars denote mean values.

As discussed in Ref. \citet{LLNS_S_k}, the LLNS equations do not
quite make sense written as a system of nonlinear SPDEs, however,
they can be linearized to obtain a well-defined linear system whose
equilibrium solutions are Gaussian fields with known covariances.
We use a \emph{finite-volume} discretization, in which space is discretized
into $N_{c}$ identical \emph{macro cells $\mathcal{V}_{j}$} of volume
$V_{c}$, and the value $\V{U}_{j}$ stored in cell $1\leq j\leq N_{c}$
is the average of the corresponding variable over the cell\begin{equation}
\V{U}_{j}(t)=\frac{1}{V_{c}}\int_{\mathcal{V}_{j}}\V{U}(\V{r},t)d\V{r}=\frac{1}{V_{c}}\int_{\mathcal{V}_{j}}\widetilde{\V{U}}(\V{r},t)d\V{r},\label{U_j_finite_volume}\end{equation}
where $\widetilde{\V{U}}$ is defined in Eq. (\ref{U_r_t_def}). Time
is discretized with a time step $\D{t}_{C}$, approximating $\V{U}(\V{r},t)$
pointwise in time with $\V{U}^{n}=\left\{ \V{U}_{1}^{n},...,\V{U}_{N_{c}}^{n}\right\} $,\[
\V{U}_{j}^{n}\approx\V{U}_{j}(n\D{t}_{C}),\]
where $n\geq0$ enumerates the macroscopic time steps. While not strictly
necessary, we will assume that each macro cell consists of an integer
number of micro cells (along each dimension of the grid), and similarly
each macro time step consists of an integer number $n_{ex}$ of micro
time steps, $\D{t}_{C}=n_{ex}\D{t}_{P}$.

In addition to the cell averages $\V{U}_{j}^{n}$, the continuum solver
needs to store the continuum normal flux $\V{F}_{j,j^{\prime}}^{n}$
through each interface $I=\mathcal{V}_{j}\cap\mathcal{V}_{j^{\prime}}$
between touching macro cells $j$ and $j^{\prime}$ during a given
time step,\begin{equation}
\V{U}_{j}^{n+1}=\V{U}_{j}^{n}-\frac{\D{t}}{V_{c}}\sum_{j^{\prime}}S_{j,j^{\prime}}\V{F}_{j,j^{\prime}}^{n},\label{U_np1_conservative}\end{equation}
where $S_{j,j^{\prime}}$ is the surface area of the interface, and
$\V{F}_{j,j^{\prime}}^{n}=-\V{F}_{j^{\prime},j}^{n}$. Here we will
absorb the various prefactors into a total transport (surface and
time integrated flux) through a given macro cell interface $I$, $\V{\Phi}_{I}^{n}=V_{c}^{-1}\D{t}S_{I}\V{F}_{I}^{n},$
which simply measures the total mass, momentum and energy transported
through the surface $I$ during the time interval from time $t$ to
time $t+\D{t}$. We arbitrarily assign one of the two possible orientations
(direction of the normal vector) for each cell-cell interface. How
the (integrated) fluxes $\V{\Phi}_{I}^{n}$ are calculated from $\V{U}_{j}^{n}$
does not formally matter; all that the hybrid method uses to advance
the solution for one macro time step are $\V{U}_{j}^{n}$, $\V{U}_{j}^{n+1}$
and $\V{\Phi}_{I}^{n}$. Therefore, any explicit conservative finite-volume
method can be substituted trivially. Given this generality, we do
not describe in any detail the numerical method used to integrate
the LLNS equations; readers can consult Ref. \citet{LLNS_S_k} for
further information.

\subsection{\label{SectionCouplingBasics}Coupling between particle and continuum
subdomains}

The hybrid method we use is based on domain decomposition,and is inspired
by Adaptive Mesh Refinement (AMR) methodology for conservation laws
\citet{BergerAMR89,AMR_Hyperbolic3D}. Our coupling scheme closely
follows previously-developed methodology for coupling a traditional
DSMC gas to a continuum fluid, first proposed in the deterministic
setting in Ref. \citet{AMAR_DSMC} and then extended to a fluctuating
continuum method in Ref. \citet{FluctuatingHydro_AMAR}. The key new
ingredient is the special handling of the collisional momentum and
energy transport across cell interfaces, not found in traditional
DSMC. For completeness, we describe the coupling algorithm in detail,
including components already described in the literature.

We split the whole computational domain into \emph{particle} and \emph{continuum}
subdomains, which communicate with each other through information
near the particle-continuum interface $I$, assumed here to be oriented
such that the flux $\V{\Phi}_{I}$ measures the transport of conserved
quantities from the particle to the continuum regions. In AMR implementations
subdomains are usually logically rectangular \emph{patches}; in our
implementation, we simply label each macro cell as either a particle
cell or a continuum cell based on whatever criterion is appropriate,
without any further restrictions on the shape or number of the resulting
subdomains. For complex fluids applications, macro cells near beads
(suspended solute) and sometimes near complex boundaries will be labeled
as particle cells. The continuum solver is completely oblivious to
what happens inside the particle subdomain and thus it need not know
how to deal with complex moving boundaries and suspended objects.
Instead, the continuum solution feels the influence of boundaries
and beads through its coupling with the particle subdomains.

The dynamic coupling between particle and continuum subdomains is
best viewed as a mutual exchange of boundary conditions between the
two regions. Broadly speaking, domain-decomposition coupling schemes
can be categorized based on the type of boundary conditions each subdomain
specifies for the other \citet{CouplingAnalysis_Weiqing}. Our scheme
is closest to a \emph{state-flux} coupling scheme based on the classification
proposed in Ref. \citet{CouplingAnalysis_Weiqing} for incompressible
solvers (the term {}``velocity-flux'' is used there since velocity
is the only state variable). A \emph{state-flux} coupling scheme is
one in which the continuum solver provides to the particle solver
the conserved variables $\V{U}$ in the continuum \emph{reservoir
macro cells} near the particle-continuum interface $I$, that is,
the continuum state is imposed as a boundary condition on the particle
region. The particle solver provides to the continuum solver the flux
$\V{\Phi}_{I}$ through the interface $I$, that is, the particle
flux is imposed as a boundary condition on the continuum subdomain.
This aims to achieve continuity of both state variables and fluxes
across the interface, and ensures strict conservation, thus making
the coupling rather robust. Note that state/flux information is only
exchanged between the continuum and particle subdomains every $n_{ex}$
particle (micro) time steps, at the beginning/end of a macro time
step. A more detailed description of the algorithm is given in Section
\ref{SectionCoupling}.

\subsubsection{Comparison with other coupling schemes}

There are several hybrid methods in the literature coupling a particle
method, and in particular, molecular dynamics (MD), with a continuum
fluid solver \citet{HybridMethods_Review}. There are two main types
of applications of such hybrids \citet{HMM_Fluids}. The first type
are problems where the particle description is localized to a region
of space where the continuum description fails, such as, for example,
a complex boundary, flow near a corner, a contact line, a drop pinchoff
region, etc.. The second type are problems where the continuum method
needs some transport coefficients, e.g., stress-strain relations,
that are not known a priori and are obtained via localized MD computations.
In the majority of existing methods a stationary solution is sought
\citet{DSMCHybrid_Boyd}, and a deterministic incompressible or isothermal
formulation of the Navier-Stokes equations is used in the continuum
\citet{MD_NS_Robbins,HMM_Fluids}. By contrast, we are interested
in a fully dynamic bidirectional coupling capable of capturing the
full range of hydrodynamic effects including sound and energy transport.
We also wish to minimize the size of the particle regions and only
localize the particle computations near suspended objects, making
it important to minimize the artifacts at the interface. As we will
demonstrate in this work, including thermal fluctuations in the continuum
formulation is necessary to obtain a proper coupling under these demanding
constraints.

The only other work we are aware of that develops a coupling between
a fluctuating compressible continuum solver and a particle method,
specifically molecular dynamics, is a coupling scheme developed over
the last several years by Coveney, Flekkoy, de Fabritiis, Delgado-Buscallioni
and collaborators \citet{FluctuatingHydroMD_Coveney,FluctuatingHydroHybrid_MD,TripleScale_Rafael}.
There are two important differences between their method and our algorithm.
Firstly, their method is (primarily, but not entirely) a flux-flux
coupling scheme, unlike our state-flux coupling. Secondly, we do not
use MD but rather I-DSMC, which, as discussed below, significantly
simplifies the handling of the continuum-particle interface.

It is not difficult to impose boundary conditions for the continuum
subdomain based on the particle data, and any consistent boundary
condition for the PDE being solved can in principle be imposed. When
the continuum solver is deterministic the fluctuations (often inappropriately
referred to as {}``noise'') in the particle data need to be filtered
using some sort of spatial and temporal averaging \citet{DSMCHybrid_Boyd},
or the continuum solver needs to be robust \citet{AMAR_DSMC,AMAR_DSMC_SAMRAI}.

The difficult part in coupling schemes is the handling of the boundary
conditions for the particle subdomain. It is very difficult to truncate
a fluid particle region without introducing artifacts in the structure
of the particle fluid, and the transport coefficients of the particle
fluid (e.g., the pressure and viscosity) critically depend on the
fluid structure. In most molecular simulations periodic boundary conditions
are used to avoid such artifacts, however, this is not possible in
our case. In order to minimize artifacts at the boundary of the particle
subdomain, most coupling schemes add an \emph{overlap region} in addition
to the reservoir region that we described above. In the overlap region,
particles are simulated even though the region belongs to the continuum
subdomain. The \emph{structure} of the fluid in the overlap region
is left to adjust to that in the particle subdomain, thus minimizing
the artifacts. At the same time, the \emph{dynamics} of the particles
in the overlap region is somehow constrained to match the underlying
continuum subdomain dynamics using, for example, various forms of
constrained Langevin-type thermostats or (non-holonomic) constrained
MD \citet{MD_NS_Robbins,MD_NSF_Robbins}. To prevent particles from
flying outside of the particle subdomain, various artificial constraining
forces are added in the reservoir region, and typically some particle
insertion/deletion scheme is used as well. The details can be very
different and tricky, and we are not aware of any detailed comparison
or even rudimentary mathematical analysis of the different types of
coupling other than the stability analysis of four idealized coupling
schemes presented in Ref. \citet{CouplingAnalysis_Weiqing}.

We do not use an overlap region in the coupling scheme we present
here. Firstly, and most importantly, because the I-DSMC fluid is structureless,
an overlap region and the associated algorithmic complications are
simply not required. The addition of an overlap region (added fluid
mass) almost always introduces some delay and errors in the coupling,
and it is usually assumed that this surface effect is negligible when
the size of the overlap region is small compared to the particle region
and when the hydro time steps are much larger than the particle time
step. These are assumptions that we do not wish to make as we try
to minimize the size of the particle subdomain.

The simple and direct coupling scheme we have presented, however,
does not work if a structured stochastic particle fluid is used, such
as, for example, a structured stochastic fluid like the Stochastic
Hard-Sphere Dynamics (SHSD) fluid \citet{SHSD}. The SHSD fluid is
weakly structured, so that at least particle insertion/deletion in
the reservoir region is not a problem. However, an overlap region
is necessary to smoothly match the fluid structure at the particle-continuum
interface. Constraining the dynamics in the overlap region can be
done in I-DSMC by introducing additional one-particle collisions (dissipation)
that scatter the particle velocities so that their mean equals the
continuum field values. However, it is not clear how to do this consistently
when fluctuations are included in the continuum solver, and in this
paper we restrict our focus on structureless (ideal) stochastic fluids.

\section{\label{SectionCoupling}Details of the Coupling Algorithm}

The basic ideas behind the state-flux coupling were already described
in Section \ref{SectionCouplingBasics}. In this section we describe
in detail the two components of the particle-continuum coupling method,
namely, the imposition of the continuum state as a boundary condition
for the particle subdomain and the imposition of the particle flux
as a boundary condition for the continuum subdomain. At the same time,
we will make clear that our coupling is not purely of the state-flux
form. The handling of the continuum subdomain is essentially unchanged
from the pure continuum case, with the only difference being the inclusion
of a refluxing step. The handling of the particle subdomain is more
complex and explained in greater detail, including pseudocodes for
several steps involved in taking a micro time step, including insertion
of reservoir particles and the tracking of the particle fluxes.

\subsection{State exchange}

\begin{figure}[tbph]
\begin{centering}
\includegraphics[width=0.5\textwidth]{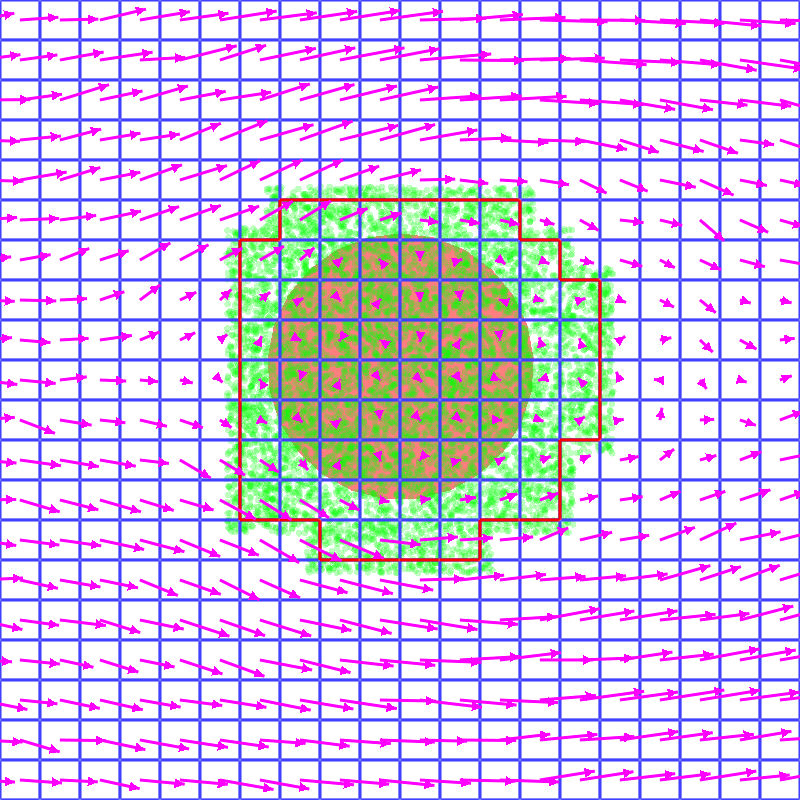}
\par\end{centering}

\caption{\label{CouplingIllustration}Illustration of a hybrid simulation of
two-dimensional plug flow around a \emph{permeable} stationary disk
(red). The macroscopic grid is shown (dark-blue lines), with each
macro cell composed of $6\times6$ micro cells (not shown). The particle
subdomain surrounds the disk and the particle-continuum interface
is shown (red). A snapshot of the I-DSMC particles is also shown (green),
including the reservoir particles outside of the particle subdomain.
The time-averaged velocity in each continuum cell is shown, revealing
the familiar plug flow velocity field that is smooth across the interface.
This example clearly demonstrates that the continuum solver feels
the stationary disk through the particle subdomain even though the
continuum solver is completely oblivious to the existence of the disk.}

\end{figure}

We first explain how the state in the \emph{reservoir} macro cells
bordering the particle subdomain, denoted by $\V{U}_{H}^{(B)}$, is
used by the particle algorithm. The micro cells that are inside the
reservoir macro cells and are sufficiently close to the particle subdomain
to affect it during a time-interval $\D{t}_{P}$ are labeled as \emph{reservoir
micro cells}. For I-DSMC fluids, assuming the length of the micro
cells along each dimension is $\V{L}_{c}\gtrapprox D$, the micro
cells immediately bordering the particle subdomain as well as all
of their neighboring micro cells need to be included in the reservoir
region. An illustration of the particle and reservoir regions is given
in Figs. \ref{HybridIllustration} and \ref{CouplingIllustration}.

At the beginning of each particle time step \emph{reservoir} \emph{particles}
are inserted randomly into the reservoir micro cells. The number of
particles inserted is based on the target density in the corresponding
reservoir macro cell. The velocities of the particles are chosen from
a Maxwell-Boltzmann or Chapman-Enskog \citet{ChapmanEnskogBook} distribution
(see discussion in Section \ref{Section_Insertion}) with mean velocity
and temperature, and also their gradients if the Chapman-Enskog distribution
is used, chosen to match the momentum and energy densities in the
corresponding macro cell. The positions of the reservoir particles
are chosen randomly uniformly (i.e., sampled from a Poisson spatial
distribution) inside the reservoir cells which does not introduce
any artifacts in the fluid structure next to the interface because
the I-DSMC fluid is ideal and structureless. This is a major advantage
of ideal fluids over more realistic structured non-ideal fluids. Note
that in Ref. \citet{DSMC_AED} we used a particle reservoir to implement
open boundary conditions in pure particle simulations, the difference
here is that the state in the reservoir cells comes from a continuum
solver instead of a pre-specified stationary flow solution.

The reservoir particles are treated just like the rest of the particles
for the duration of a particle time step. First they are advectively
propagated (streamed) along with all the other particles, and the
total mass, momentum and energy transported by particles advectively
through the particle-continuum interface, $\V{\Phi}_{P}^{(I)}$, is
recorded. Particles that, at the end of the time step, are not in
either a reservoir micro cell or in the particle subdomain are discarded,
and then stochastic collisions are processed between the remaining
particles. In the traditional DSMC algorithm, collisions occur only
between particles inside the same micro cell, and thus all of the
particles outside the particle subdomain can be discarded \citet{AMAR_DSMC,FluctuatingHydro_AMAR}.
However, in the I-DSMC algorithm particles in neighboring cells may
also collide, and thus the reservoir particles must be kept until
the end of the time step. Collisions between a particle in the particle
subdomain and a particle in a reservoir cell lead to collisional exchange
of momentum and energy through the interface as well and this contribution
must also be included in $\V{\Phi}_{P}^{(I)}$. Note that the reservoir
particles at the very edge of the reservoir region do not have an
isotropic particle environment and thus there are artifacts in the
collisions which they experience, however, this does not matter since
it is only essential that the particles in the particle subdomain
not feel the presence of the interface.

\subsection{\label{SectionReFlux}Flux exchange}

After the particle subdomain is advanced for $n_{ex}$ (micro) time
steps, the particle flux $\V{\Phi}_{P}^{(I)}$ is imposed as a boundary
condition in the continuum solver so that it can complete its (macro)
time step. This flux exchange ensures strict conservation, which is
essential for long-time stability. Assume that the continuum solver
is a one-step explicit method that uses only stencils of width one,
that is, calculating the flux for a given macro cell-cell interface
only uses the values in the adjacent cells. Under such a scenario,
the continuum solver only needs to calculate fluxes for the cell-cell
interfaces between continuum cells, and once the particle flux $\V{\Phi}_{P}^{(I)}$
is known the continuum time step (\ref{U_np1_conservative}) can be
completed. It is obvious that in this simplified scenario the coupling
is purely of the state-flux form. However, in practice we use a method
that combines pieces of state and flux exchange between the particle
and continuum regions.

First, the particle state is partly used to advance the continuum
solver to the next time step. Our continuum solver is a multi-stage
method and uses stencils of width two in each stage, thus using an
effective stencils that can be significantly larger than one cell
wide \citet{LLNS_S_k}. While one can imagine modifying the continuum
solver to use specialized boundary stencils near the particle-continuum
interface (e.g., one-sided differencing or extrapolation), this is
not only more complex to implement but it is also less accurate. Instead,
the continuum method solves for the fields over the whole computational
domain (continuum patch in AMAR terminology) and uses hydrodynamic
values for the particle subdomain (particle patch) obtained from the
particle solver. These values are then used to calculate \emph{provisional}
fluxes $\V{\Phi}_{H}^{(I)}$ and take a \emph{provisional} time step,
as if there were no particle subdomain. This makes the implementation
of the continuum solver essentially oblivious to the existence of
the particle regions, however, it does require the particle solver
to provide reasonable conserved values for all of the macro particle
cells. This may not be possible for cells where the continuum hydrodynamic
description itself breaks down, for example, cells that overlap with
or are completely covered by impermeable beads or features of a complex
boundary. In such \emph{empty cells} the best that can be done is
to provide reasonable hydrodynamic values, for example, values based
on the steady state compatible with the specified macroscopic boundary
conditions. For partially empty cells, that is, macro cells that are
only partially obscured, one can use the uncovered fraction of the
hydro cell to estimate hydrodynamic values for the whole cell. In
practice, we have found that as long as empty cells are sufficiently
far from the particle-continuum interface (in particular, empty cells
must not border the continuum subdomain) the exact improvised hydrodynamic
values do not matter much. Note that for permeable beads there is
no problem with empty cells since the fluid covers the whole domain.

Secondly, the provisional continuum fluxes are partly used to advance
the particle subdomain to the same point in time as the continuum
solver. Specifically, a linear interpolation between the current continuum
state and the provisional state is used as a boundary condition for
the reservoir particles. This temporal interpolation is expected to
improve the temporal accuracy of the coupling, although we are not
aware of any detailed analysis. Note that if $n_{ex}=1$ this interpolation
makes no difference and the provisional continuum fluxes are never
actually used.

Once the particle solver advances $n_{ex}$ time steps, a particle
flux $\V{\Phi}_{P}^{(I)}$ is available and it is imposed in the continuum
solver to finalize the provisional time step. Specifically, hydrodynamic
values in the particle macro cells are overwritten based on the actual
particle state, ignoring the provisional prediction. In order to correct
the provisional fluxes, a \emph{refluxing} procedure is used in which
the state $\V{U}_{H}^{(B)}$ in each of the continuum cells that border
the particle-continuum interface are changed to reflect the particle
$\V{\Phi}_{P}^{(I)}$ rather than the provisional flux $\V{\Phi}_{H}^{(I)}$,\[
\V{U}_{H}^{(B)}\leftarrow\V{U}_{H}^{(B)}-\V{\Phi}_{H}^{(I)}+\V{\Phi}_{P}^{(I)}.\]
This refluxing step ensures strict conservation and ensures continuity
of the fluxes across the interface, in addition to continuity of the
state.

\subsection{Taking a macro time step}

Algorithm \ref{MacroTimeStep} summarizes the hybrid algorithm and
the steps involved in advancing both the simulation time by one macro
time step $\D{t}_{C}=n_{ex}\D{t}_{P}$. Note that at the beginning
of the simulation, we initialize the hydrodynamic values for the continuum
solver, consistent with the particle data in the particle subdomain
and generated randomly from the known (Gaussian) equilibrium distributions
in the continuum subdomain.

\begin{Algorithm}{\label{MacroTimeStep}Take a macro time step by
updating the continuum state $\V{U}_{H}$ from time $t$ to time $t+\D{t}_{C}$.}

\begin{enumerate}
\item \emph{\label{TakeProvisionalStep}Provisionally advance the continuum
solver}: Compute a \emph{provisional} macro solution $\V{U}_{H}^{next}$
at time $t+\D{t}_{C}$ everywhere, including the particle subdomain,
with an estimated (integrated) provisional flux $\V{\Phi}_{H}^{(I)}$
through the particle-continuum interface. Reset the particle flux
$\V{\Phi}_{P}^{(I)}\leftarrow\V{0}$.
\item \emph{Advance the particle solver}: Take $n_{ex}$ micro time steps
(see Algorithm \ref{DSMCTimeStep}):

\begin{enumerate}
\item At the beginning of each particle time step reservoir particles are
inserted at the boundary of the particle subdomain with positions
and velocities based on a linear interpolation between $\V{U}_{H}$
and $\V{U}_{H}^{next}$ (see Algorithm \ref{InsertReservoirParticles}).
This is how the continuum state is imposed as a boundary condition
on the particle subdomain.
\item All particles are propagated advectively by $\D{t}_{P}$ and stochastic
collisions are processed, accumulating a particle flux $\V{\Phi}_{P}^{(I)}$
(see Algorithm \ref{DSMCFlux}).
\end{enumerate}
\item \emph{Synchronize} the continuum and particle solutions\emph{:}

\begin{enumerate}
\item \emph{Advance}: Accept the provisional macro state, $\V{U}_{H}\leftarrow\V{U}_{H}^{next}$.
\item \emph{Correct}: The continuum solution in the particle subdomain $\V{U}_{H}^{(P)}$
is replaced with cell averages of the particle state at time $t+\D{t}_{C}$,
thus forming a composite state over the whole domain.
\item \emph{Reflux: }The continuum solution $\V{U}_{H}^{(B)}$ in the macro
cells bordering the particle subdomain is corrected based on the particle
flux, $\V{U}_{H}^{(B)}\leftarrow\V{U}_{H}^{(B)}-\V{\Phi}_{H}^{(I)}+\V{\Phi}_{P}^{(I)}$.
This effectively imposes the particle flux as a boundary condition
on the continuum and ensures conservation in the hybrid update.
\item \emph{Update} the partitioning between particle and continuum cells
if necessary. Note that this step may convert a continuum cell into
an \emph{unfilled} (devoid of particles) particle cell.
\end{enumerate}
\end{enumerate}
\end{Algorithm}

\subsection{\label{SectionTimeStepping}Taking a micro time step}

Taking a micro time step is described in Algorithm \ref{DSMCTimeStep},
and the remaining subsections give further details on the two most
important procedures used. The initial particle configuration can
most easily be generated by marking all macro cells in the particle
domain as unfilled.

\begin{Algorithm}{\label{DSMCTimeStep}Take an I-DSMC time step.
We do not include details about the handling of the non-DSMC (solute)
particles here. Note that the micro time step counter $n_{P}$ should
be re-initialized to zero after every $n_{ex}$ time steps.}

\begin{enumerate}
\item Visit all macro cells that overlap the reservoir region or are unfilled
(i.e., recently converted from continuum to particle) one by one,
and insert trial particles in each of them based on the continuum
state $\V{U}_{H}$, as described in Algorithm \ref{InsertReservoirParticles}.
\item Update the clock $t\leftarrow t+\D{t}_{P}$ and advance the particle
subdomain step counter $n_{P}\leftarrow n_{P}+1$. Note that when
an event-driven algorithm is used this may involve processing any
number of events that occur over the time interval $\D{t}_{P}$ \citet{DSMC_AED}.
\item Move all I-DSMC particles to the present time, updating the total
kinetic mass, momentum and energy flux through the coupling interface
$\V{F}_{P}^{(I)}$ whenever a particle crosses from the particle into
the continuum subdomain or vice versa, as detailed in Algorithm \ref{DSMCFlux}.
\item Perform stochastic collisions between fluid particles, including the
particles in the reservoir region. Keep track of the total \emph{collisional}
momentum and energy flux through the coupling interface by accounting
for the amount of momentum $\D{\V{p}}_{ij}=m\D{\V{\upsilon}}_{ij}$
and kinetic energy $\D{e}_{ij}$ transferred from a regular particle
$i$ (in the particle subdomain) to a reservoir particle $j$ (in
the continuum subdomain), $\V{\Phi}_{P}^{(I)}\leftarrow\V{\Phi}_{P}^{(I)}+\left[0,\D{\V{p}}_{ij},\D{e}_{ij}\right]$
.
\item Remove all particles from the continuum subdomain.
\item Linearly interpolate the continuum state to the present time, \[
\V{U}_{H}\leftarrow\frac{\left(n_{ex}-n_{P}\right)\V{U}_{H}+\V{U}_{H}^{next}}{n_{ex}-n_{P}+1}\]

\end{enumerate}
\end{Algorithm}

\subsubsection{Particle flux tracking}

As particles are advected during a particle time step, they may cross
from a particle to a continuum cell and vice versa, and we need to
keep track of the resulting fluxes. Note that a particle may cross
up to $d$ cell interfaces near corners, where $d$ is the spatial
dimension, and may even recross the same interface twice near hard-wall
boundaries. Therefore, ray tracing is the most simple and reliable
way to account for all particle fluxes correctly. For the majority
of the particles in the interior, far from corners and hard walls,
the usual quick DSMC update will, however, be sufficient.

\begin{Algorithm}{\label{DSMCFlux}Move the fluid particle $i$ from
time $t$ to time $t+\D{t}_{P}$ and determine whether it crosses
the coupling interface. We will assume that during a particle time
step no particle can move more than one macro cell length along each
dimension (in practice particles typically move only a fraction of
a micro cell).}

\begin{enumerate}
\item Store information on the macro cell $c_{old}$ to which the particle
belongs, and then tentatively update the position of the particle
$\V{r}_{i}\leftarrow\V{r}_{i}+\V{\upsilon}_{i}\D{t}_{P}$, and find
the tentative macro cell $c_{i}$ and micro cell $b_{i}$ to which
the particle moves, taking into account periodic boundary conditions.
\item If $c_{old}$ and $c_{i}$ are near a boundary, go to step \ref{RayTracePath}.
If $c_{old}\equiv c_{i}$, or if $c_{old}$ and $c_{i}$ are both
continuum or are both particle cells and at least one of them is not
at a corner, accept the new particle position and go to step \ref{AcceptNewPosition}.
\item \label{RayTracePath}Undo the tentative particle update, $\V{r}_{i}\leftarrow\V{r}_{i}-\V{\upsilon}_{i}\D{t}_{P}$,
and then ray trace the path of the particle during this time step
from one macro cell-cell interface $I_{c}$ to the next, accounting
for boundary conditions (e.g., wrapping around periodic boundaries
and colliding the particle with any hard walls it encounters \citet{DSMC_AED}).
Every time the particle crosses from a particle to a continuum cell,
update the particle flux at the cell interface $I_{c}$, $\V{\Phi}_{P}^{(I_{c})}\leftarrow\V{\Phi}_{P}^{(I_{c})}+\left[m,m\V{\upsilon}_{i},m\upsilon_{i}^{2}/2\right]$,
similarly, if the particle crosses from a continuum to a particle
cell, $\V{\Phi}_{P}^{(I_{c})}\leftarrow\V{\Phi}_{P}^{(I_{c})}-\left[m,m\V{\upsilon}_{i},m\upsilon_{i}^{2}/2\right]$.
\item \label{AcceptNewPosition}If the new particle micro cell $b_{i}$
is neither in the particle subdomain nor in the reservoir region,
remove the particle from the system.
\end{enumerate}
\end{Algorithm}

\subsubsection{\label{Section_Insertion}Inserting reservoir particles}

At every particle time step, \emph{reservoir particles} need to be
inserted into the reservoir region or unfilled continuum cells. These
particles may later enter the particle subdomain or they may be discarded,
while the trial particles in unfilled cells will be retained unless
they leave the particle subdomain (see Algorithm \ref{DSMCFlux}).
When inserting particles in an unfilled macro cell, it is important
to maintain strict momentum and energy conservation by ensuring that
the inserted particles have exactly the same total momentum and energy
as the previous continuum values. This avoids global drifts of momentum
and energy, which will be important in several of the examples we
present in Section \ref{sec:Results}. It is not possible to ensure
strict mass conservation because of quantization effects, but by using
smart (randomized) rounding one can avoid any spurious drifts in the
mass.

The velocity distribution for the trial particles should, at first
approximation, be chosen from a Maxwell-Boltzmann distribution. However,
it is well-known from kinetic theory that the presence of shear and
temperature gradients skews the distribution, specifically, to first
order in the gradients the Chapman-Enskog distribution is obtained
\citet{ChapmanEnskogBook,ChapmanEnskog_Generation}. It is important
to note that only the \emph{kinetic} contribution to the viscosity
enters in the Chapman-Enskog distribution and not the full viscosity
which also includes a \emph{collisional} viscosity for dense gases
\citet{ChapmanEnskogBook}. Previous work on deterministic DSMC hybrids
has, as expected, found that using the Chapman-Enskog distribution
improves the accuracy of the hybrids \citet{AMAR_DSMC,DSMCHybrid_Boyd}.
However, in the presence of transient fluctuations and the associated
transient gradients it becomes less clear what the appropriate distribution
to use is, as we observe numerically in Section \ref{SectionMismatch}.

Certainly at pure equilibrium we know that the Maxwell-Boltzmann distribution
is correct despite the presence of fluctuating gradients. At the same
time, we know that the CE distribution ought to be used in the presence
of constant macroscopic gradients, as it is required in order to obtain
the correct kinetic contribution to the viscous stress tensor \citet{ChapmanEnskogFluxes}.
The inability to estimate time-dependent mean gradients from just
a single fluctuating realization forces the use of instantaneous gradients,
which are unreliable due to the statistical uncertainty and can become
unphysically large when $N_{0}$ is small (say $N_{0}<50$). In some
cases the background macroscopic gradients may be known \emph{a priori}
(for example, Couette or Fourier flows) and they can be used without
trying to numerically estimate them, otherwise, they can be assumed
to be zero or numerically estimated by performing some spatio-temporal
averaging \citet{FluctuatingHydro_AMAR}. Note that here we assume
that the particles are uniformly distributed (Poisson spatial distribution)
as in an ideal gas, and we do not try to take the density gradient
into account (but see the procedure described in the Appendix of Ref.
\citet{FluctuatingHydro_AMAR}).

\begin{Algorithm}{\label{InsertReservoirParticles}Insert trial particles
in the reservoir or unfilled macro cell $c$ with centroid $\V{r}_{c}$,
taking into account the target density $\rho_{c}$, velocity $\V{v}_{c}$
and temperature $T_{c}$ in cell $c$, calculated from the conserved
cell state $\V{U}_{c}$. If using the Chapman-Enskog distribution,
also take into account the estimated local shear rate $\grad_{c}\V{v}$
and local temperature gradient $\grad_{c}T$.}

\begin{enumerate}
\item Build a list $\mathcal{L}_{rc}$ of the $N_{rc}$ micro cells contained
in the macro cell $c$ that need to be filled with particles. This
typically excludes micro cells that are partially covered by impermeable
beads or boundaries so as to avoid generating overlaps.
\item Determine the total number $N_{p}$ of trial particles to insert into
the reservoir portion of $c$ by sampling from the binomial distribution
\[
P(N_{p})=\left(\begin{array}{c}
\bar{N}_{p}\\
N_{p}\end{array}\right)p^{N_{p}}\left(1-p\right)^{\bar{N}_{p}-N_{p}},\]
where $\bar{N}_{p}=\left\lfloor \rho_{c}V_{c}/m\right\rfloor $ is
the total expected number of particles in $c$, $V_{c}$ is the volume
of $c$, and $p=N_{rc}/N_{sc}$, where $N_{sc}$ is the number of
micro subcells per macro cell. For sufficiently large $\bar{N}_{p}$
this can be well-approximated by a Gaussian distribution, which can
be sampled faster.
\item For each of the $N_{p}$ trial particles to be inserted, do:

\begin{enumerate}
\item Choose a micro cell $b$ uniformly from the $N_{rc}$ micro cells
in the list $\mathcal{L}_{rc}$.
\item Generate a random particle position $\V{r}_{i}\leftarrow\V{r}_{c}+\V{r}_{rel}$
uniformly inside micro cell $b$.
\item Generate a random relative velocity for the particle $\V{v}_{rel}$
from the Maxwell-Boltzmann or Chapman-Enskog distribution \citet{ChapmanEnskog_Generation},
and set the particle velocity by taking into account the desired continuum
state in cell $c$ and, if available, its estimated gradient, $\V{\upsilon}_{i}\leftarrow\V{v}_{c}+\left(\grad_{c}\V{v}\right)\V{r}_{rel}+\V{v}_{rel}$.
\item If the cell $c$ is an \emph{unfilled} macro cell, keep track of the
total momentum $\V{P}$ and energy $E$ of the $N_{p}$ trial particles,
to be adjusted in Step \ref{CorrectUnfilledCell} for conservation.
\end{enumerate}
\item \label{CorrectUnfilledCell}If cell $c$ was unfilled and $N_{p}>1$,
then correct the particle velocities to match the desired total momentum
$\V{P}_{c}=\V{p}_{c}V_{c}$ and energy $E_{c}=V_{c}e_{c}$ inside
macro cell $c$, thus ensuring exact conservation:

\begin{enumerate}
\item Calculate the scaling factor\[
\alpha^{2}=\frac{E_{c}-P_{c}^{2}/(2mN_{p})}{E-P^{2}/(2mN_{p})}\]
 and velocity shift $\D{\V{\upsilon}}=\left(\V{P}-\V{P}_{c}\right)/\left(\alpha mN_{p}\right)$.
\item Scale and shift the velocity for every trial particle $i$, $\V{\upsilon}_{i}\leftarrow\alpha\left(\V{\upsilon}_{i}-\D{\V{\upsilon}}\right)$. 
\end{enumerate}
\end{enumerate}
\end{Algorithm}

\section{\label{sec:Results}Results}

In this section we provide extensive tests of the hybrid scheme, in
both equilibrium and non-equilibrium situations, and in both two and
three dimensions. Our goal is to access how well the hybrid method
can reproduce results obtained with a pure particle method, which
we consider the gold standard.

We have implemented our hybrid method in a code that can handle both
two and three dimensional systems. Of course, one can study one- and
two-dimensional flows with the three dimensional code by using periodic
boundaries along the remaining dimensions. We refer to this as quasi
one- or two-dimensional simulations. At the same time, both (I-)DSMC
and the LLNS equations have a truly two-dimensional formulation, which
we have also implemented for testing purposes. Transport coefficients
for two-dimensional particle models formally diverge in the infinite-time
limit (see the discussion in Ref. \citet{VACF_2Divergence}) and it
is not obvious that the Navier-Stokes equations are a proper coarse-graining
of the microscopic dynamics. However, this divergence is very slow
(logarithmic) and it will be mollified (bounded) by finite system
size, and we will therefore not need to concern ourselves with these
issues. In the first three examples, we use the three-dimensional
particle and continuum codes, and use the two-dimensional code only
for the adiabatic piston example for computational reasons.

We have also implemented continuum solvers for both the full non-linear
and the \emph{linearized} LLNS equations. As discussed in more detail
in Ref. \citet{LLNS_S_k}, the nonlinear LLNS equations are mathematically
ill-defined and this can lead to breakdown in the numerical solution
such as negative densities or temperatures. At the same time, the
linearized equations are not able to describe a wide range of physical
phenomena such as the effect of fluctuations on the mean flow; they
also omit a number of terms of order $N_{0}^{-1}$, where $N_{0}$
is the average number of particles per continuum cell (e.g., the center-of-mass
kinetic energy). If the number of particles per continuum cell is
sufficiently large (in our experience, $N_{0}>75$) the fluctuations
are small and the difference between the linear and nonlinear hydrodynamic
solvers is very small, and we prefer to use the nonlinear solver.
We will use $N_{0}\sim100$ in our hybrid simulations.

In our implementation, the continuum solver can either be the more
accurate third-order Runge-Kutta (RK3) temporal integrator developed
in Ref. \citet{LLNS_S_k} or a more classical stochastic MacCormack
integrator \citet{FluctuatingHydro_Garcia}. The analysis in Ref.
\citet{LLNS_S_k} shows that obtaining reasonably-accurate equilibrium
fluctuations with predictor-corrector methods for the diffusive fluxes,
as used in the MacCormack scheme, requires using a continuum time
step $\D{t}_{C}$ that is a fraction of the CFL stability limit $\D{t}_{CFL}$.
In the simulations we present here we have typically used a time step
$\D{t}\approx0.2\D{t}_{CFL}$, which is typically still about $5$
times larger than the particle time step $\D{t}_{P}$, and we have
found little impact of the exact value of $\D{t}_{C}$.

The hybrid method requires estimates of the transport coefficients
of the particle fluid, notably the viscosity and the thermal conductivity.
For traditional DSMC at low densities there are rather accurate theoretical
estimates of the viscosity and thermal conductivity \citet{DSMC_CellSizeError,DSMC_TimeStepError2},
however, it is nontrivial to obtain reasonably accurate theoretical
values at the higher densities we use in I-DSMC because of the importance
of multi-particle correlations. We therefore estimate the transport
coefficients numerically. For this purpose we simulate a system with
periodic boundaries along two of the directions and isothermal stick
wall boundaries along the other direction. To estimate the viscosity
we apply a shear flow by moving one of the wall boundaries at constant
speed inducing a Couette flow with an approximately constant shear
gradient $\bar{\grad}\V{v}=\frac{1}{2}(\grad\V{v}+\grad\V{v}^{T})$.
We then calculate the steady-state stress tensor $\M{\sigma}$ by
averaging over all particles $i$ and colliding pairs of particles
$ij$ over a long time interval $\D{t}$,\[
\M{\sigma}=\M{\sigma}_{k}+\mathbf{\M{\sigma}}_{c}=m\left\langle \V{\upsilon}_{i}\otimes\V{\upsilon}_{i}\right\rangle +\frac{\left\langle \V{r}_{ij}\otimes\D{\V{p}_{ij}}\right\rangle _{c}}{\D{t}},\]
where $\M{\sigma}_{k}=P\M{I}+2\eta_{k}\bar{\grad}\V{v}$ is the kinetic
contribution giving the \emph{kinetic viscosity} $\eta_{k}$, and
$\mathbf{\M{\sigma}}_{c}=2\eta_{c}\bar{\grad}\V{v}$ is the collisional
contribution giving the \emph{collisional viscosity} $\eta_{c}$,
$\eta=\eta_{k}+\eta_{c}$. We exclude particles that are close to
the wall boundaries when calculating these averages to minimize finite-size
effects. Similarly, for the thermal conductivity we apply a small
constant temperature gradient $\grad T$ by setting the two walls
at different temperatures, and we also impose the required density
gradient to maintain mechanical equilibrium (constant pressure). We
then calculate the steady-state heat flux vector $\M{\xi}=\mu\grad T$,\[
\M{\xi}=\M{\xi}_{k}+\M{\xi}_{c}=m\left\langle \frac{\upsilon_{i}^{2}}{2}\V{\upsilon}_{i}\right\rangle +\frac{\left\langle \left(\D{e}_{ij}\right)\V{r}_{ij}\right\rangle _{c}}{\D{t}},\]
from which we obtain the kinetic and collisional contributions to
the thermal conductivity. There are alternative methods that one can
use to calculate the transport coefficients, using both equilibrium
and non-equilibrium settings, however, we have found the above method
to be most accurate for a given computational effort if only moderate
accuracy is desired. Results from different non-equilibrium methods
are found to be within $5-10\%$ of each other.

\subsection{\label{SectionMismatch}Mismatch at the interface}

Previous work has studied a hybrid scheme very similar to the one
described here for several quasi one-dimensional situations \citet{FluctuatingHydro_AMAR}.
Reference \citet{FluctuatingHydro_AMAR} first studied a pure equilibrium
situation in which one part of a periodic domain was covered by a
particle subdomain, and found that the stochastic hybrid scheme was
able to reproduce the spatio-temporal correlations in equilibrium
fluctuations very well, although some mismatch at the particle/continuum
interface was found. Here we explore this mismatch more carefully
by studying a quasi one-dimensional periodic system where the middle
portion of the domain is filled with particles and the rest is continuum.
In each macro cell, we compute the average density, temperature and
velocity and their variances with high accuracy.

We first calculate the mean conserved quantities (mean density, momentum
density, and energy density) in each macro cell, and then calculate
the mean velocity and temperature from those, for example, $\left\langle v\right\rangle =\left\langle j\right\rangle /\left\langle \rho\right\rangle $
instead of averaging the instantaneous velocities, $\left\langle v\right\rangle =\left\langle j/\rho\right\rangle $.
As shown in Ref. \citet{UnbiasedEstimates_Garcia}, the approach we
use leads to an unbiased estimate of the mean, while the latter has
a bias when there are correlations between the fluctuations of the
different hydrodynamic variables. The variances are estimated from
the instantaneous values, e.g., $\left\langle \delta v^{2}\right\rangle =\left\langle \left(j/\rho\right)^{2}\right\rangle -\left\langle j/\rho\right\rangle ^{2}$.
It is possible to construct unbiased estimates for the variances as
well \citet{UnbiasedEstimates_Garcia}, however, this is somewhat
more involved and the bias is rather small compared to the artifacts
we are focusing on.

\begin{figure}[tbph]
\begin{centering}
\subfigure[]{\label{InterfaceMismatch_means}\includegraphics[width=0.65\textwidth]{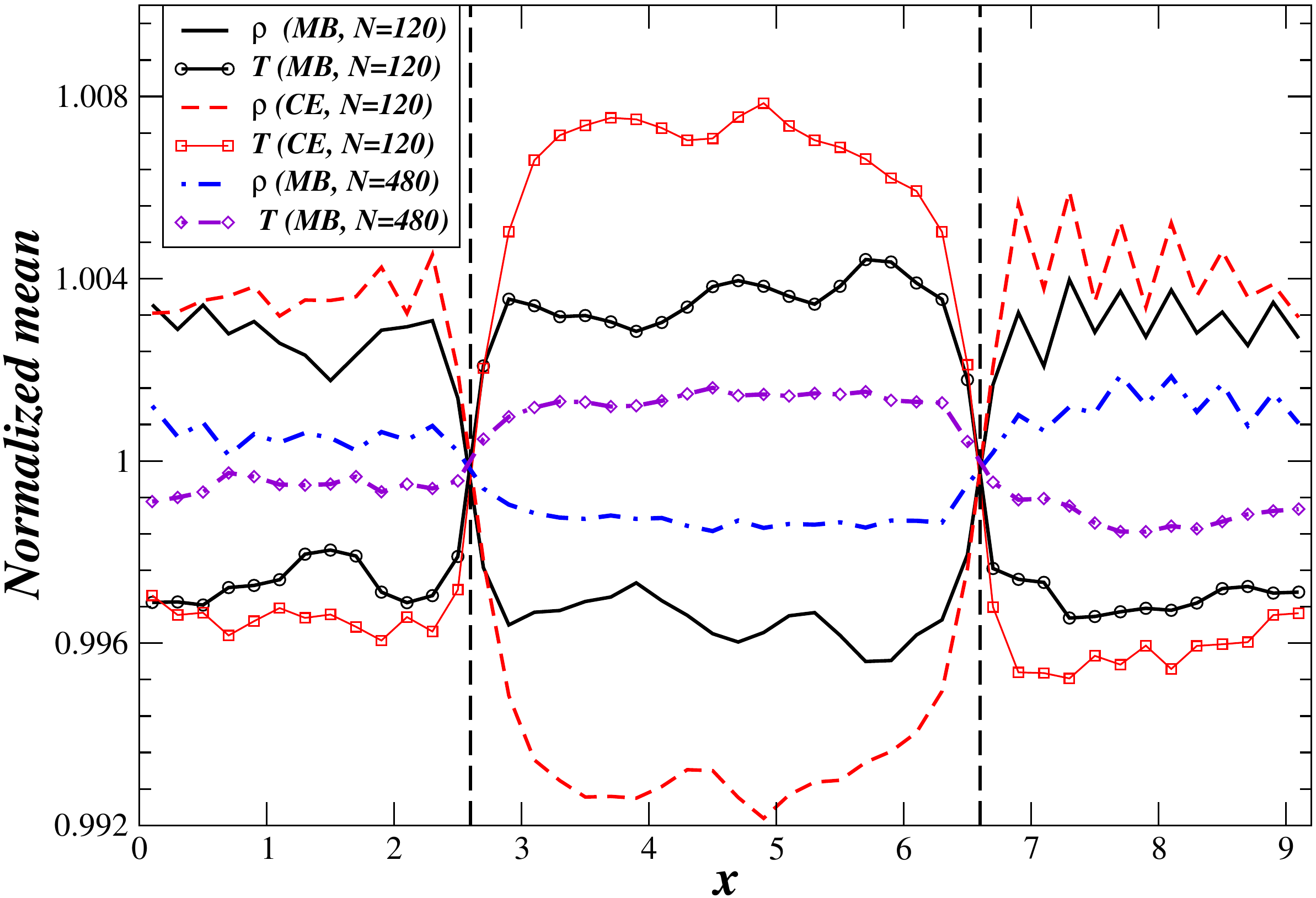}}
\par\end{centering}

\begin{centering}
\subfigure[]{\label{InterfaceMismatch_variances}\includegraphics[width=0.65\textwidth]{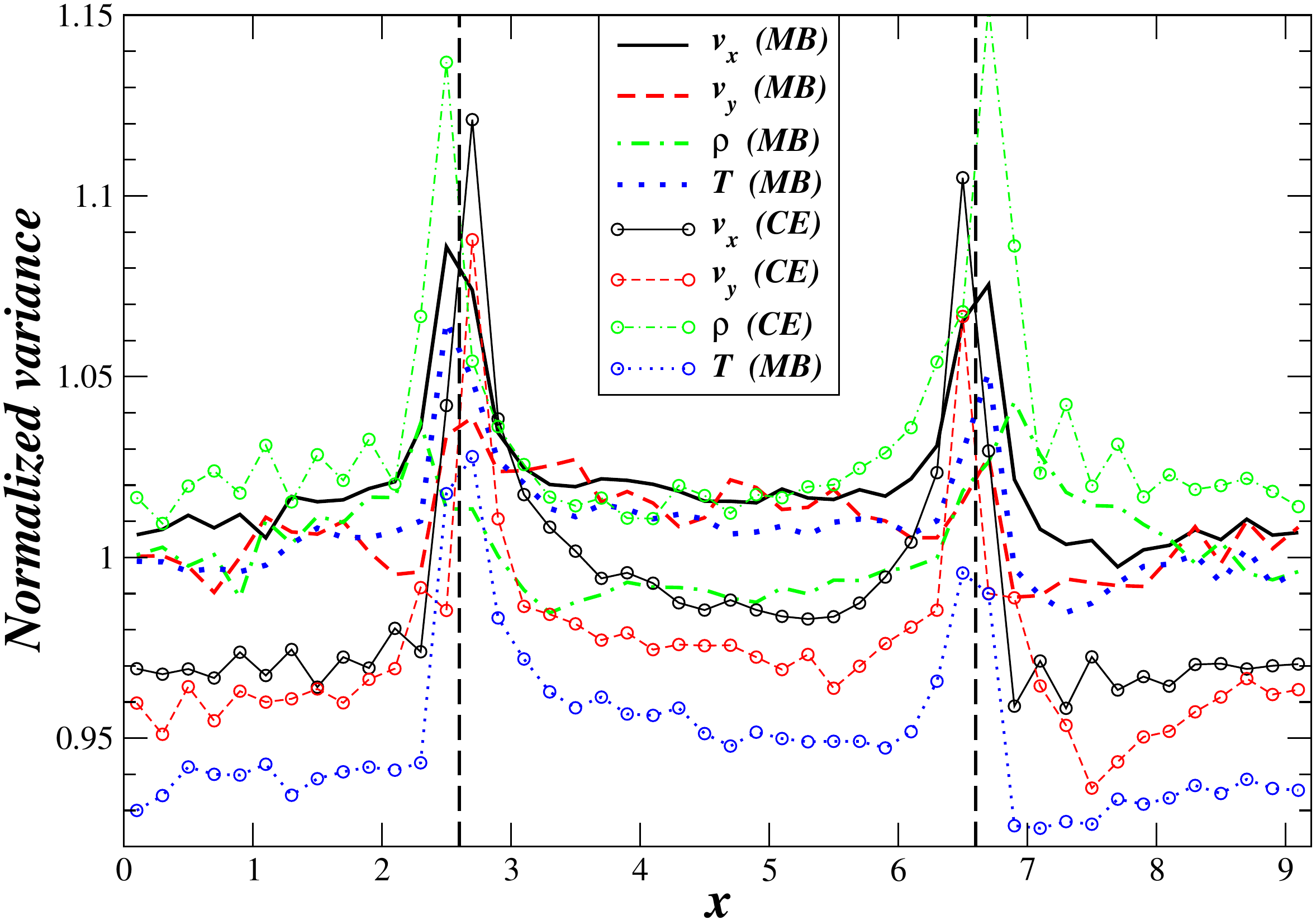}}
\par\end{centering}

\caption{\label{fig:InterfaceMismatch}Normalized means and variances of the
hydrodynamic variables in each of the $46$ macroscopic cells of a
quasi one-dimensional periodic system where the middle portion ($20$
continuum cells, in-between dashed vertical lines) is the particle
subdomain. The velocities of the reservoir particles are either samples
from a Maxwell-Boltzmann (MB, lines only) or a Chapman-Enskog distribution
(CE, lines and symbols). \subref{InterfaceMismatch_means} Unbiased
\citet{UnbiasedEstimates_Garcia} estimates of the mean density and
temperature (we averaged about $2.5\cdot10^{6}$ samples taken every
macro step) for two different sizes of the continuum cells, ones containing
$N_{0}=120$ particles on average, and ones containing $N_{0}=480$
particles. Only the MB results are shown for the larger cells for
clarity, with similar results observed for CE. \subref{InterfaceMismatch_variances}
Estimates of the variances of the hydrodynamic variables for $N_{0}=120$
particles per continuum cell (also an average over $2.5\cdot10^{6}$
samples).}

\end{figure}

In Fig. \ref{fig:InterfaceMismatch} we show the means and variances
along the length of the system, normalized by the expected values
\citet{FluctuatingHydro_Garcia}. For velocity, the mean is zero to
within statistical uncertainty in both the particle and continuum
subdomains, and we do not show it in the figure. However, a small
mismatch is clearly seen in Fig. \ref{InterfaceMismatch_means} between
the density and temperature in the particle and continuum subdomains.
The mismatch is such that the pressure $\left\langle p\right\rangle =\left\langle \rho\right\rangle R\left\langle T\right\rangle $
is constant across the interface, that is, the particle and continuum
subdomains are in mechanical equilibrium but not in true thermodynamic
equilibrium. In Appendix \ref{AppendixFluxMismatch} we show that
this kind of mismatch is expected because the average particle fluxes
coming from the reservoir particles inserted in the continuum subdomains
have a bias of order $N_{0}^{-1}$, where $N_{0}$ is the average
number of particles per macroscopic cell. Because our coupling matches
both the state and the fluxes across the interface this bias makes
it impossible for the particle and continuum to reach true thermodynamic
equilibrium. The theory in Appendix \ref{AppendixFluxMismatch} suggests
that the size of the mismatch is of order $N_{0}^{-1}$, consistent
with the results in Fig. \ref{InterfaceMismatch_means}; however,
the crude theoretical estimates do not actually give the steady state
since they assume equilibrium to begin with. That the cells near the
interface are not in equilibrium is reflected in the variances of
the hydrodynamic variables, which have a spike near the particle-continuum
interface, as shown in Fig. \ref{InterfaceMismatch_variances}. We
have observed that the relative magnitude of this spike does not depend
on $N_{0}$.

The cause of the mismatch is the fact that we use the instantaneous
values of the local density, velocity and temperature when generating
the velocities for the reservoir particles. This is necessary because
we cannot in general obtain estimates of the time-dependent mean values
from running a single realization, and are forced to use the instantaneous
values. One can use some sort of spatio-temporal averaging to obtain
estimates of the local means, however, this introduces additional
time and length scales into the algorithm that do not have an obvious
physical interpretation. For steady-state problems it may be possible
to use running means to avoid the mismatch we observe, however, this
is not possible in a general dynamic context. Deeper theoretical understanding
of the connection between the microscopic dynamics and the LLNS equations
is necessary to design a more consistent approach.

Another complex issue that arises in the fluctuating hybrid is whether
to use the Maxwell-Boltzmann (MB) or the Chapman-Enskog (CE) distributions
when generating the velocities of the reservoir particles (see discussion
in Section \ref{Section_Insertion}). In Fig. \ref{InterfaceMismatch_means}
we compare the size of the mismatch at the interface when the MB and
CE distributions are used. For the CE distribution we obtained a local
estimate of the gradient using simple centered differences of the
instantaneous hydrodynamic variables. As seen in the figure, there
is a greater discrepancy when the CE distribution is used, especially
for the variances. We will therefore adopt a compromise in which we
use the MB distribution for all calculations reported here. In cases
when there is a macroscopic background gradient that is specified
\emph{a priori} (e.g., shear flows) we can use that gradient in the
CE distribution.

Note that we have performed numerous additional quasi-one dimensional
tests of the coupling that we do not describe here for brevity. For
example, we have tested the matching of the shear stress tensor in
a shear flow parallel to the particle-continuum interface by verifying
that a linear velocity profile is obtained without any slope discontinuity
at the interface (see Appendix D.2 in Ref. \citet{FluctuatingHydroHybrid_MD}).
We have also reproduced the results reported in Ref. \citet{FluctuatingHydro_AMAR},
such as the presence of unphysical long-range correlations in the
fluctuations in the particle region when the deterministic instead
of the stochastic hybrid is used.

\subsection{\label{Section_S_kw}Dynamic Structure Factors}

The hydrodynamics of the spontaneous thermal fluctuations in the I-DSMC
fluid is expected to be described by the Landau-Lifshitz Navier-Stokes
(LLNS) equations for the fluctuating field $\V{U}=(\rho_{0}+\delta\rho,\delta v,T_{0}+\delta T)$
linearized around a reference equilibrium state $\V{U}_{0}=(\rho_{0},\V{v}_{0}=\V{0},T_{0})$
\citet{Landau:Fluid,FluctHydroNonEq_Book,LLNS_S_k}. By solving these
equations in the Fourier wavevector-frequency $(\V{k},\omega)$ domain
for $\widehat{\V{U}}(\V{k},\omega)=(\widehat{\delta\rho},\widehat{\delta v},\widehat{\delta T})$
and performing an ensemble average over the fluctuating stresses we
can obtain the equilibrium (stationary) spatio-temporal correlations
(covariance) of the fluctuating fields. We express these correlations
in terms of the $3\times3$ symmetric positive-definite \emph{hydrodynamic
structure factor matrix} $\M{S}_{H}(\V{k},\omega)=\left\langle \widehat{\V{U}}\widehat{\V{U}}^{\star}\right\rangle $
\citet{LLNS_S_k}, which is essentially the spatio-temporal spectrum
of the fluctuating fields. Integrating $\M{S}_{H}(\V{k},\omega)$
over frequencies gives the \emph{hydrostatic structure factor matrix}
$\M{S}_{H}(\V{k})$, which turns out to be diagonal since in any given
snapshot the hydrodynamic variables are uncorrelated at equilibrium.

We non-dimensionalize $\M{S}_{H}(\V{k},\omega)$ so that $\M{S}_{H}(\V{k})$
is the identity matrix. For example, the density-density correlations
are given by the dimensionless structure factor\[
S_{\rho}(\V{k},\omega)=\left(\rho_{0}c_{0}^{-2}k_{B}T_{0}\right)^{-1}\left\langle \hat{\rho}(\V{k},\omega)\hat{\rho}^{\star}(\V{k},\omega)\right\rangle ,\]
and we express the spatio-temporal cross-correlation between density
and velocity through the dimensionless structure factor\[
S_{\rho,v}(\V{k},\omega)=\left(\rho_{0}c_{0}^{-2}k_{B}T_{0}\right)^{-\frac{1}{2}}\left(\rho_{0}^{-1}k_{B}T_{0}\right)^{-\frac{1}{2}}\left\langle \hat{\rho}(\V{k},\omega)\hat{v}^{\star}(\V{k},\omega)\right\rangle ,\]
where $c_{0}^{2}=k_{B}T_{0}/m$ is the isothermal speed of sound.
Reference \citet{FluctuatingHydro_AMAR} demonstrated that a hybrid
method very similar to the one we described here correctly reproduces
the density-density time-correlation function $S_{\rho}(\V{k},t)$
for large wavelengths in a quasi one-dimensional periodic system.
The density-density dynamic structure factor $S_{\rho}(\V{k},\omega)$
is often the only one considered because it is accessible experimentally
via light scattering measurements and thus most familiar \citet{FluctHydroNonEq_Book}.
The full dynamic structure factor matrix $\M{S}_{H}(\V{k},\omega)$
is a more complete measure of the spatio-temporal evolution of the
thermal fluctuations and includes both sound (hyperbolic) and dissipative
(diffusive) effects. It is therefore important to show that the hybrid
scheme correctly reproduces $\M{S}_{H}(\V{k},\omega)$ as compared
to a purely particle simulation and demonstrate that the hybrid is
capable of capturing the propagation of spontaneous thermal fluctuations
across the particle-continuum interface.

\subsubsection{Bulk Dynamic Structure Factor}

For a bulk fluid, i.e., for periodic boundary conditions, it is well-known
\citet{FluctHydroNonEq_Book} that the density-density component $S_{\rho}(\V{k},\omega)$
and the temperature-temperature component $S_{T}(\V{k},\omega)$ of
$\M{S}_{H}(\V{k},\omega)$ exhibit three peaks for a given wavevector
$\V{k}$. There is one central Rayleigh peak at $\omega=0$ which
comes from entropic fluctuations. There is also two symmetric Brillouin
peaks at $\omega\approx c_{s}k$, where $c_{s}$ is the adiabatic
speed of sound, which come from the isoentropic propagation of sound
waves induced by the fluctuations. For the components of the velocity
parallel to the wavevector the dynamic structure factors $S_{\V{v}_{\parallel}}(\V{k},\omega)$
exhibit all three peaks, while for the component perpendicular to
the wavevector $S_{v_{\perp}}(\V{k},\omega)$ lacks the central peak.

\begin{figure}[tbph]
\begin{centering}
\includegraphics[width=0.65\textwidth]{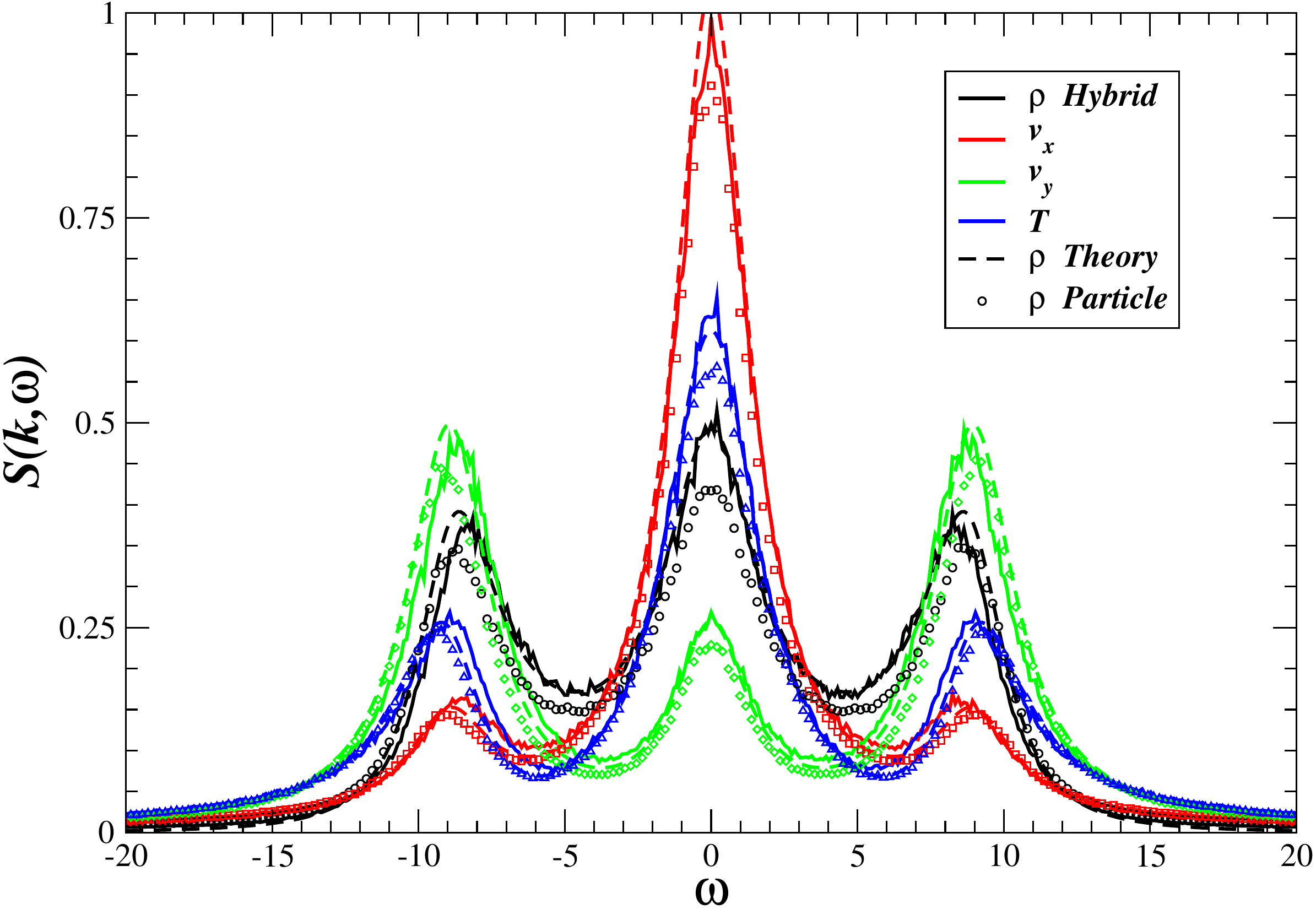}
\par\end{centering}

\begin{centering}
\includegraphics[width=0.65\textwidth]{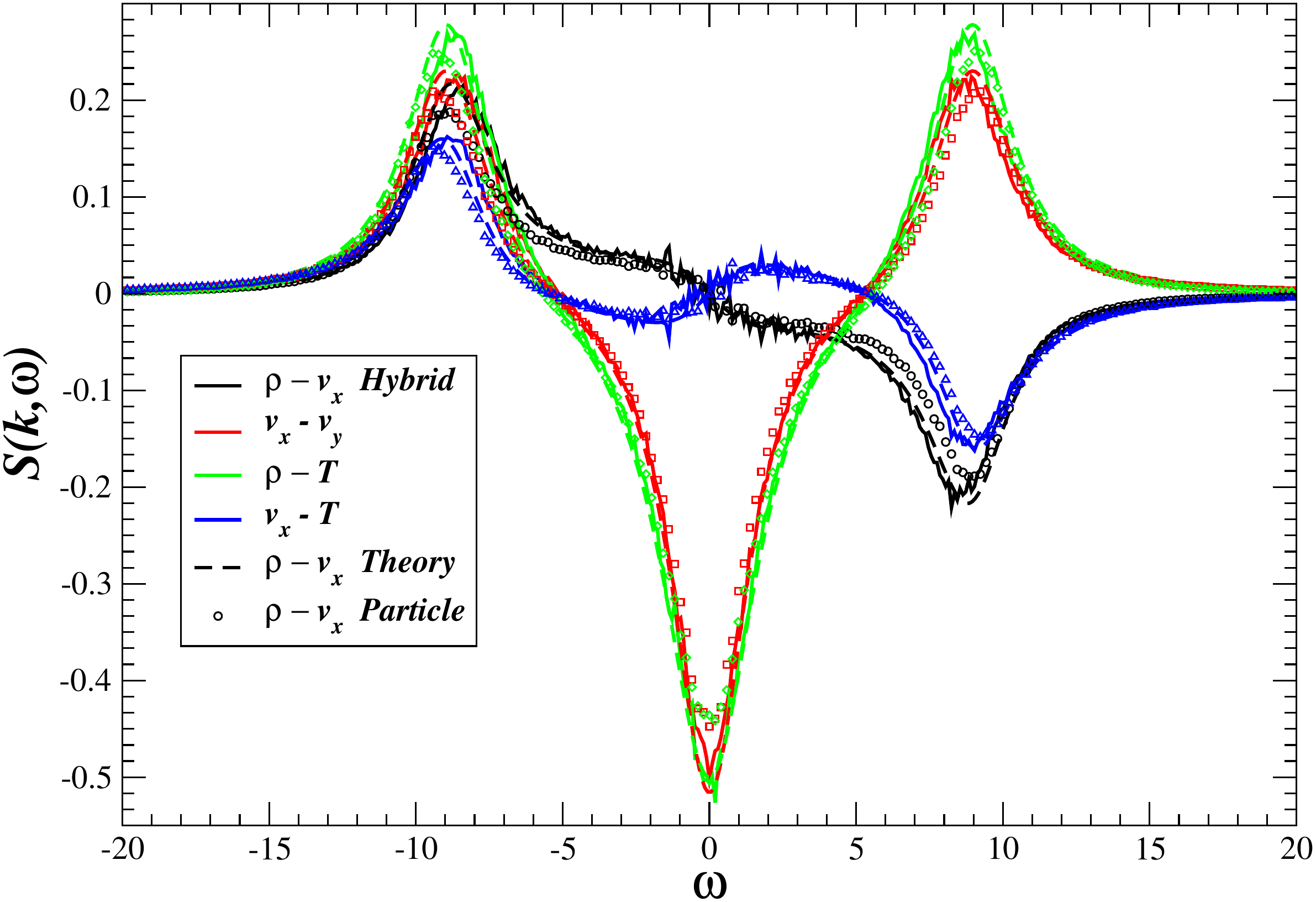}
\par\end{centering}

\caption{\label{fig:S_kw_bulk}Discrete dynamic structure factors for waveindices
$q_{x}=k_{x}L_{x}/(2\pi)=1$, $q_{y}=2$ and $q_{z}=0$ for a quasi
two-dimensional system with lengths $L_{x}=L_{y}=2$, $L_{z}=0.2$,
split into a grid of $10\times10\times1$ macro cells. Each macro
cell contains about $120$ I-DSMC particles of diameter $D=0.04$
(density $\phi=0.5$, and collision frequency $\chi=0.62$). An average
over $5$ runs each containing $10^{4}$ temporal snapshots is employed.
We perform pure particle runs and also hybrid runs in which only a
strip 4 macro cells along the $x$ axes was filled with particles
and the rest handled with a continuum solver. The different hydrodynamic
pairs of variables are shown with different colors, using a solid
line for the result from the hybrid runs, symbols for the results
of the pure particle runs, and a dashed line for the theoretical predictions
based on the linearized LLNS equations (solved using the computer
algebra system Maple). (\emph{Top}) The diagonal components $S_{\rho}(\V{k},\omega)$,
$S_{v_{x}}(\V{k},\omega)$, $S_{v_{y}}(\V{k},\omega)$ and $S_{T}(\V{k},\omega)$.
(\emph{Bottom}) The off-diagonal components (cross-correlations) $S_{\rho,v_{x}}(\V{k},\omega)$,
$S_{v_{x},v_{y}}(\V{k},\omega)$, $S_{\rho,T}(\V{k},\omega)$ and
$S_{v_{x},T}(\V{k},\omega)$.}

\end{figure}

Figure \ref{fig:S_kw_bulk} shows selected dynamic structure factors
for a quasi two-dimensional system with periodic boundary conditions.
The simulation box is composed of $10\times10\times1$ macro cells,
each cell containing about $120$ particles. Finite-volume averages
of the hydrodynamic conserved variables and the corresponding spatial
Discrete Fourier Transforms (DFTs) were then calculated for each cell
every $10$ macro time steps and a temporal DFT was used to obtain
discrete dynamic structure factors \citet{LLNS_S_k} for several wavenumbers.
In the first set of particle runs, the whole domain was filled with
particles and the macro cells were only used to sample hydrodynamic
fields. In the second set of hybrid runs, the central portion of the
simulation box was split along the $x$ axes and designated as a particle
subdomain, and the remaining two thirds of the domain were continuum.

In Fig. \ref{fig:S_kw_bulk} we show the results for a wavevector
$\V{k}$ that is neither parallel nor perpendicular to the particle-continuum
interface so as to test the propagation of both perpendicular and
tangential fluctuations across the interface. The results show very
little discrepancy between the pure particle and the hybrid runs,
and they also conform to the theoretical predictions based on the
LLNS equations. Perfect agreement is not expected because the theory
is for the spectrum of the continuum field while the numerical results
are discrete spectra of cell averages of the field, a distinction
that becomes important when the wavelength is comparable to the cell
size. Additionally, even purely continuum calculations do not reproduce
the theory exactly because of spatio-temporal discretization artifacts.

\subsubsection{Dynamic Structure Factors for Finite Systems}

The previous section discussed the {}``bulk'' dynamic structure
factors, as obtained by using periodic boundary conditions. For non-periodic
(i.e., finite) systems equilibrium statistical mechanics requires
that the static structure factor be oblivious to the presence of walls.
However, the dynamic structure factors exhibit additional peaks due
to the reflections of sound waves from the boundaries. At a hard-wall
boundary surface $\partial\Omega$ with normal vector $\V{n}$ either
Dirichlet or von Neumann boundary conditions need to be imposed on
the components of the velocity and the temperature (the boundary condition
for density follows from these two). Two particularly common types
of boundaries are:

\begin{description}
\item [{Thermal~walls}] for which a stick condition is imposed on the
velocity, $\V{v}_{\partial\Omega}=0$, and the temperature is fixed,
$T_{\partial\Omega}=T_{0}$.
\item [{Adiabatic~walls}] for which a slip condition is imposed on the
velocity, $\V{n}\cdot\grad\V{v}_{\parallel}=0$ and $v_{\perp}=0$,
and there is no heat conduction through the wall, $\V{n}\cdot\grad T=0$.
\end{description}
In particle simulations, these boundary conditions are imposed by
employing standard rules for particle reflection at the boundaries
\citet{DSMC_AED}. We describe the corresponding handling in continuum
simulations in Appendix \ref{AppendixWallsLLNS}. In Appendix \ref{Appendix_S_kw_walls}
we derive the form of the additional peaks in the dynamic structure
factor for adiabatic walls by solving the linearized LLNS equations
with the appropriate conditions. 

\begin{figure}[tbph]
\begin{centering}
\includegraphics[width=0.6\textwidth]{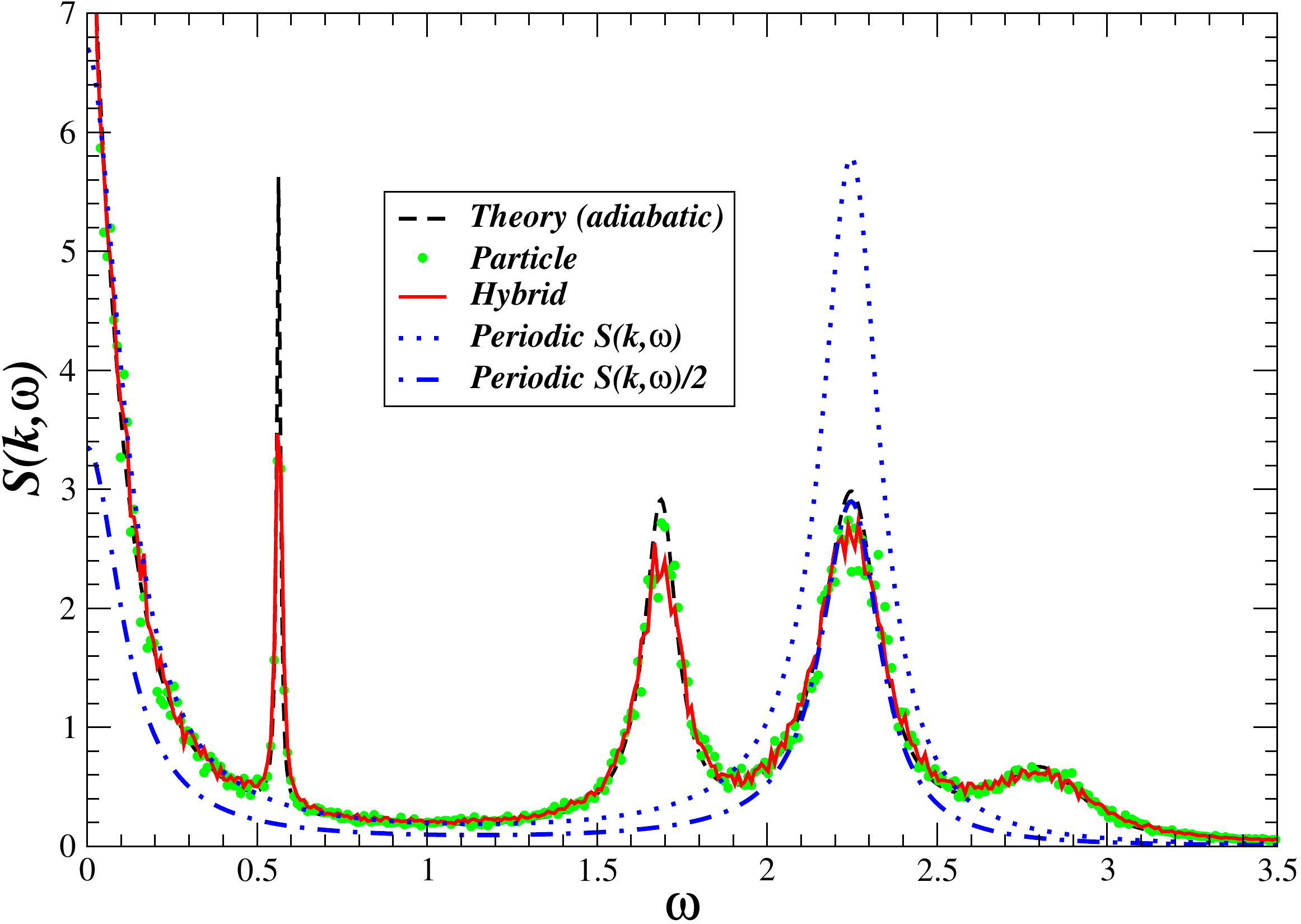}
\par\end{centering}

\begin{centering}
\includegraphics[width=0.6\textwidth]{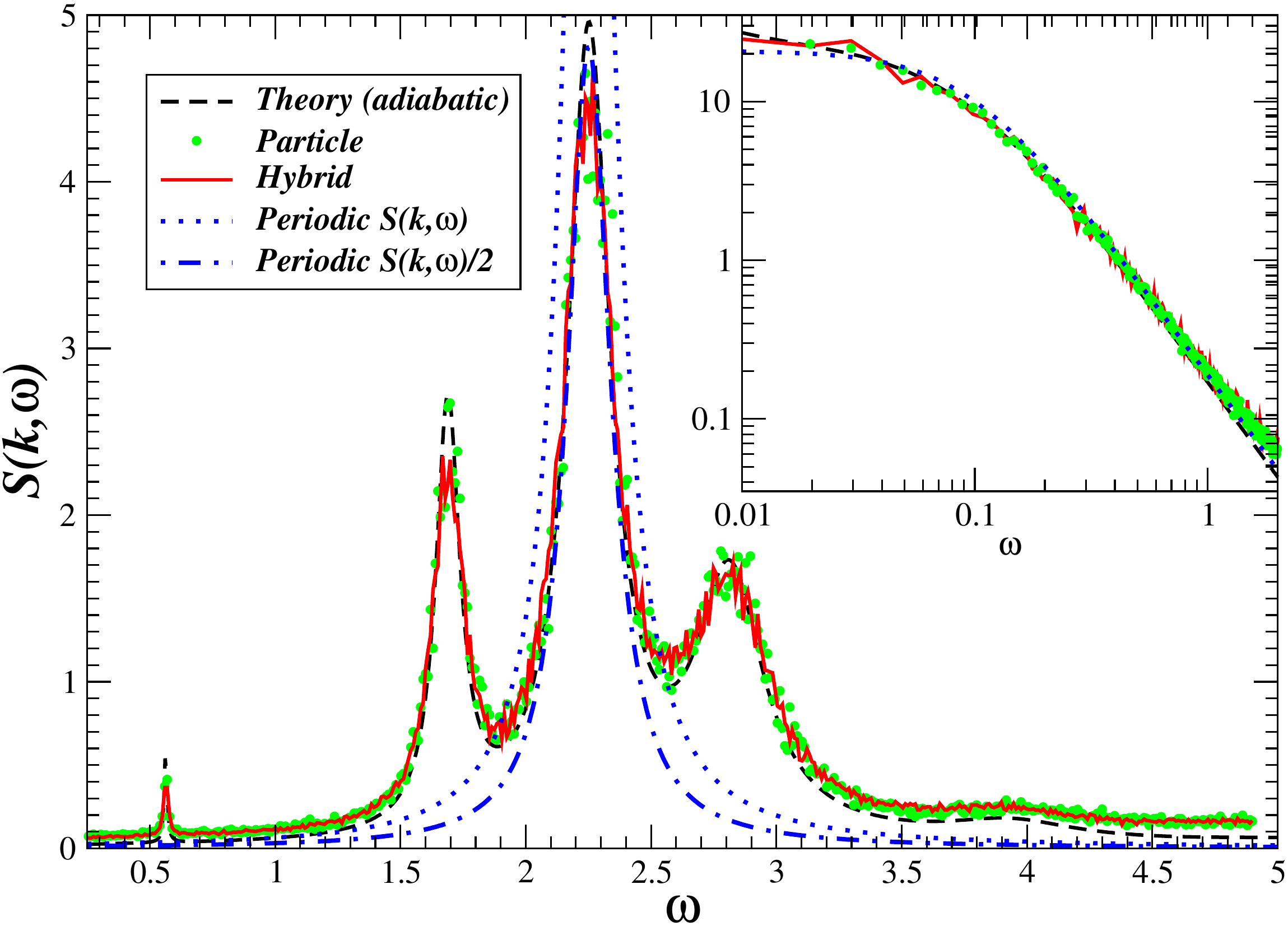}
\par\end{centering}

\caption{\label{fig:S_kw_walls}Discrete dynamic structure factors for a quasi
one-dimensional system with length $L=7.2$ (corresponding to 36 continuum
cells) bounded by adiabatic walls, for waveindex $q=2$ and wavevector
$k=2\pi q/L\approx1.75$. The I-DSMC fluid parameters are as in Fig.
\ref{fig:S_kw_bulk}. We perform pure particle runs and also hybrid
runs in which the middle third of the domain is filled with particles
and the rest handled with a continuum solver. The results from purely
particle runs are shown with symbols, while the results from the hybrid
are shown with a solid line. The theoretical predictions based on
the linearized LLNS equations and the equations in Appendix \ref{Appendix_S_kw_walls}
(solved using the computer algebra system Maple) are shown with a
dashed line for adiabatic boundaries and dotted line for periodic
boundaries. Since the magnitude of the Brillouin peaks shrinks to
one half the bulk value in the presence of adiabatic walls, we also
show the result with periodic boundaries scaled by $1/2$ (dashed-dotted
line). (\emph{Top}) Dynamic structure factor for density, $S_{\rho}(k,\omega)$,
showing the Rayleigh peak and the multiple Brillouin peaks. (\emph{Bottom})
Dynamic structure factor for the component of velocity perpendicular
to the wall, $S_{v_{\perp}}(k,\omega)$, which lacks the Rayleigh
peak. The corresponding correlations for either of the parallel velocity
components, $S_{v_{\parallel}}(k,\omega)$, have only a Rayleigh peak,
shown in the inset on a log-log scale.}

\end{figure}

In Figure \ref{fig:S_kw_walls} we show dynamic structure factors
for a quasi one-dimensional system bounded by adiabatic walls. As
for the bulk (periodic) case in the previous section, we perform both
purely particle runs and also hybrid runs in which the middle third
of the domain is designated as a particle subdomain. Additional peaks
due to the reflections of sound waves from the boundaries are clearly
visible and correctly predicted by the LLNS equations and also accurately
reproduced by both the purely continuum solution (not shown) and the
hybrid. Similar agreement (not shown) is obtained between the particle,
continuum and the hybrid runs for thermal walls. These results show
that the hybrid is capable of capturing the dynamics of the fluctuations
even in the presence of boundaries. Note that when the deterministic
hybrid scheme is used one obtains essentially the correct shape of
the peaks in the structure factor (not shown), however, the magnitude
is smaller (by a factor of $2.5$ for the example in Fig. \ref{fig:S_kw_walls})
than the correct value due to the reduced level of fluctuations.

\subsection{\label{SectionVACF}Bead VACF}

As an illustration of the correct hydrodynamic behavior of the hybrid
algorithm, we study the velocity autocorrelation function (VACF) $C(t)=\left\langle v_{x}(0)v_{x}(t)\right\rangle $
for a large \emph{neutrally-buoyant} impermeable \emph{bead} of mass
$M$ and radius $R$ diffusing through a dense Maxwell I-DSMC stochastic
fluid \citet{SHSD} of particles with mass $m\ll M$ and collision
diameter $D\ll R$ and density (volume fraction) $\phi$ {[}mass density
$\rho=6m\phi/(\pi D^{3})$]. The VACF problem is relevant to the modeling
of polymer chains or (nano)colloids in solution (i.e., \emph{complex
fluids}), in particular, the integral of the VACF determines the diffusion
coefficient which is an important macroscopic quantity. Furthermore,
the very first MD studies of the VACF for fluid molecules led to the
discovery of a long power-law tail in $C(t)$ \citet{VACF_Alder}
which has since become a standard test for hydrodynamic behavior of
methods for complex fluids \citet{BrownianLB_VACF,MPCD_VACF,FluctuatingHydro_FluidOnly,BrownianSRD_Review,BrownianParticle_SIBM,BrownianParticle_IncompressibleSmoothed,SHSD_PRL}.

The fluctuation-dissipation principle \citet{FluctuationDissipation_Kubo}
points out that $C(t)$ is exactly the decaying speed of a bead that
initially has a unit speed, if only viscous dissipation was present
without fluctuations, and the equipartition principle tells us that
$C(0)=\left\langle v_{x}^{2}\right\rangle =kT/2M$. Using these two
observations and assuming that the dissipation is well-described by
a continuum approximation with stick boundary conditions on a sphere
of radius $R_{H}$, $C(t)$ has been calculated from the linearized
(compressible) Navier-Stokes (NS) equations \citet{VACF_FluctHydro,BrownianCompressibility_Zwanzig}.
The results are analytically complex even in the Laplace domain, however,
at short times an inviscid compressible approximation applies. At
large times the compressibility does not play a role and the incompressible
NS equations can be used to predict the long-time tail \citet{BrownianCompressibility_Zwanzig,VACF_LagrangeTheory}.
At short times, $t<t_{c}=2R_{H}/c_{s}$, the major effect of compressibility
is that sound waves generated by the motion of the suspended particle
carry away a fraction of the momentum with the sound speed $c_{s}$,
so that the VACF quickly decays from its initial value $C(0)=k_{B}T/M$
to $C(t_{c})\approx k_{B}T/M_{eff}$, where $M_{eff}=M+2\pi R^{3}\rho/3$
\citet{BrownianCompressibility_Zwanzig}. At long times, $t>t_{visc}=4\rho R_{H}^{2}/3\eta$,
the VACF decays as with an asymptotic power-law tail $(k_{B}T/M)(8\sqrt{3\pi})^{-1}(t/t_{visc})^{-3/2}$,
in disagreement with predictions based on the Langevin equation (Brownian
dynamics), $C(t)=(k_{B}T/M)\exp\left(-6\pi R_{H}\eta t/M\right)$.

We performed purely particle simulations of a diffusing bead in various
I-DSMC fluids in Refs. \citet{SHSD_PRL,SHSD}. In purely particle
methods the length of the runs necessary to achieve sufficient accuracy
in the region of the hydrodynamic tail is often prohibitively large
for beads much larger than the fluid particles themselves. It is necessary
to use hybrid methods and limit the particle region to the vicinity
of the bead in order to achieve a sufficient separation of the molecular,
sonic, viscous, and diffusive time scales and study sufficiently large
box sizes over sufficiently long times. In the results we report here
we have used an impermeable hard bead for easier comparison with existing
theory; similar results are obtained using permeable beads. The interaction
between the I-DSMC fluid particles and the bead is treated as if the
fluid particles are hard spheres of diameter $D_{s}$, chosen to be
somewhat smaller than their interaction diameter with other fluid
particles (specifically, we use $D_{s}=D/4$) for computational efficiency
reasons, using an event-driven algorithm \citet{DSMC_AED}. Upon collision
with the bead the relative velocity of the fluid particle is reversed
in order to provide a no-slip condition at the surface of the suspended
sphere \citet{BrownianSRD_Review,DSMC_AED} (slip boundaries give
qualitatively identical results). We have estimated the effective
(hydrodynamic) colloid radius $R_{H}$ from numerical measurements
of the Stokes friction force $F=-6\pi R_{H}\eta v$ and found it to
be somewhat larger than the hard-core collision radius $R+D_{s}/2$,
but for the calculations below we use $R_{H}=R+D_{s}/2$. Since we
used periodic boundary conditions with a box of length $L$, there
are artifacts in $C(t)$ after about the time at which sound waves
generated by its periodic images reach the particle, $t_{L}=L/c_{s}$.
The averaging procedure that we used in order to eliminate some of
the noise from the tail does not properly resolve these sound effects%
\footnote{To calculate the VACF, we first calculated the average mean-square
displacement $\D{r}(t)$ by averaging over a long run and numerically
differentiated this to obtain a time-dependent diffusion coefficient
$D(t)$. We then smoothed $D(t)$ by fitting a quadratic polynomial
over short-time intervals spaced on a logarithmic scale (so that more
points were averaged over in the tail), and obtained the VACF by differentiating
the smoothed $D(t)$. This procedure produces well-resolved tails,
however, it obscures features over short time scales compared to the
smoothing interval. A more direct velocity autocorrelation calculation
was used at very short times.%
}, even though they are visible in the results shown below. Furthermore,
there are strong finite-size effects that manifest themselves as a
rapid decay of the VACF for times longer than the viscous time-scale
$\rho L^{2}/\eta$ \citet{BrownianParticle_SIBM}.

For the hybrid calculations, we localize the particle subdomain to
the continuum cells that overlap or are close to the moving bead.
The location of the particle subdomain is updated periodically as
the bead moves. The algorithm that we use tries to fit the particle
subdomain as closely around the bead but ensuring that there are a
certain number of micro cells in-between the surface of the bead and
the particle-continuum interface. The exact shape of the particle
subdomain thus changes as the bead moves and the number of particles
employed by the hybrid fluctuates, especially when the bead is small
compared to the continuum cells. Obtaining reasonably-accurate results
for the VACF at long times requires very long runs. We found that
it is crucial to strictly conserve momentum in the hybrid when unfilled
continuum cells are transformed into particle cells. Otherwise very
slow drifts in the momentum of the system appear due to the use of
periodic boundary conditions, and this drift changes the tail of $C(t)$,
especially for massive beads where the typical bead speed is already
small compared to the typical fluid particle speed. We used the MacCormack
solver and the linearized formulation of the LLNS equations for these
simulations, however, similar results are obtained with the non-linear
Runge-Kutta solver as long as the macro cells are sufficiently large.
As discussed earlier, the use of the linearized formulation makes
it possible to reduce the size of the continuum cells without introducing
numerical problems due to the non-linearity of the equations.

\begin{figure}[tbph]
\begin{centering}
\subfigure[]{\label{VACF.hybrid.small}\includegraphics[width=0.49\columnwidth]{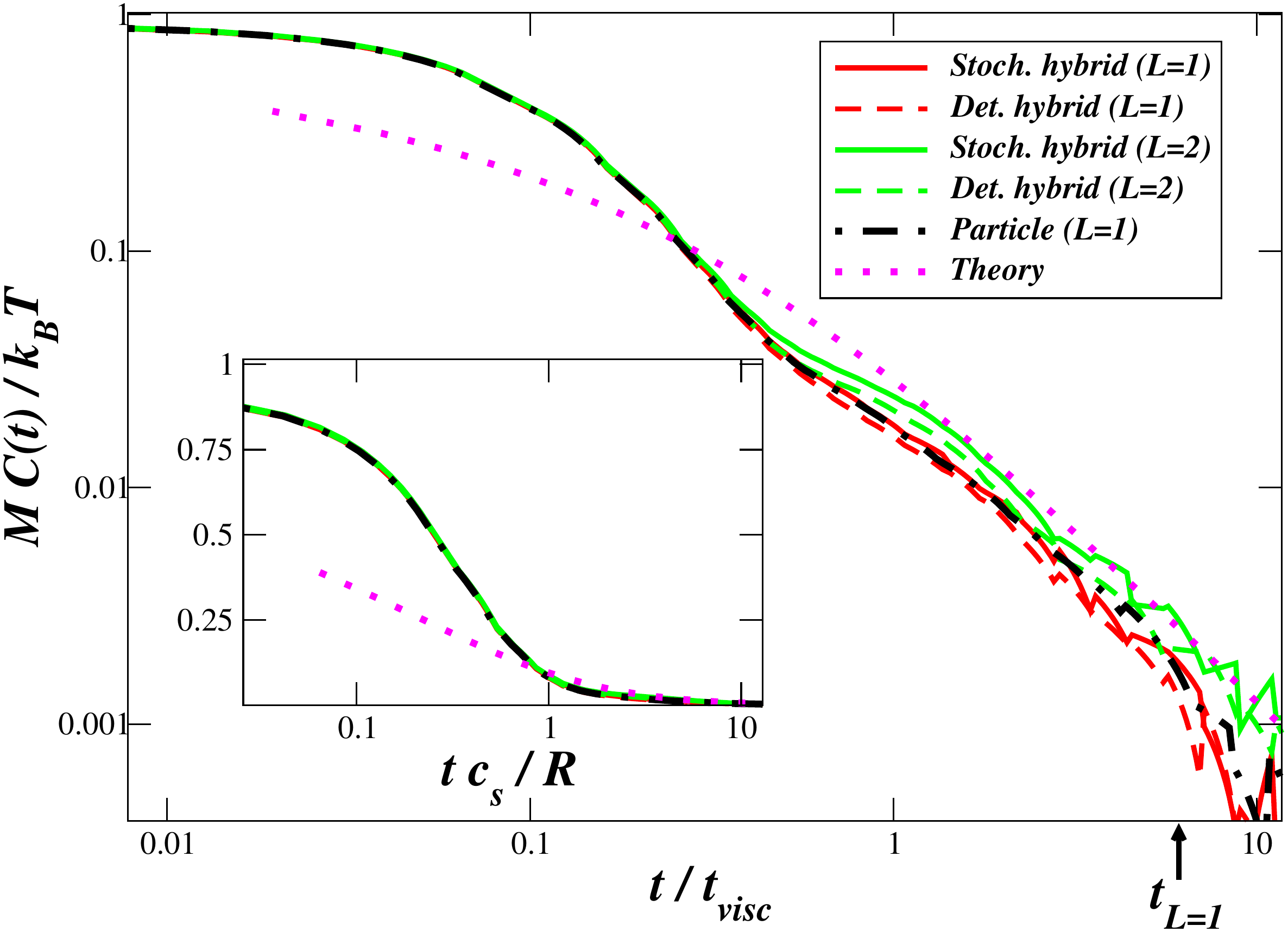}}\subfigure[]{\label{VACF.hybrid.huge}\includegraphics[width=0.49\columnwidth]{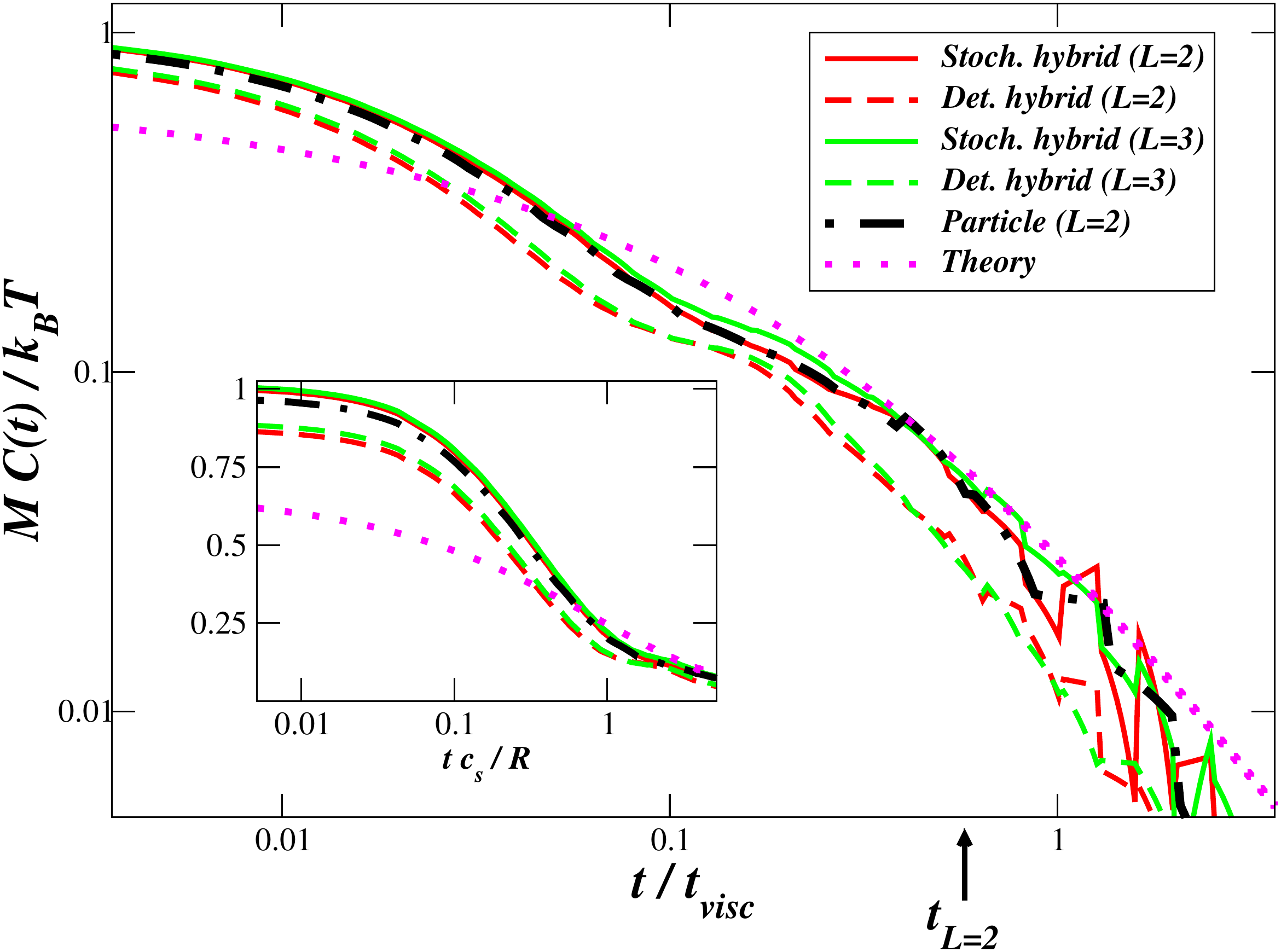}}
\par\end{centering}

\caption{\label{VACF.hybrid}Normalized velocity autocorrelation function (VACF)
$C(t)/\left(k_{B}T/M\right)$ for a neutrally-buoyant hard spherical
bead of mass $M$ suspended in a fluid of I-DSMC particles of diameter
$D$, for two different bead sizes, a small bead of radius $R=1.25D$
\subref{VACF.hybrid.small} and a large bead of radius $R=6.25D$
\subref{VACF.hybrid.huge}. A log-log scale is used to emphasize the
long-time power law tail and the time is normalized by the viscous
time $t_{visc}$, so that the results should be approximately independent
of the actual bead radius. The inset shows the initial decay of the
VACF on a semi-log scale, where the time is normalized by the sonic
time scale $t_{c}$. Periodic boundary conditions with a cubic cell
of length $L$ are employed, and the sound crossing time $t_{L}$
is indicated. Results from purely particle simulations are shown with
a dashed-dotted line, and the incompressible hydrodynamic theory is
shown in with a dotted line. Results from hybrid runs are also shown
with a solid line for the stochastic hybrid and dashed line for the
deterministic hybrid, for both the same box size as the particle runs
(red) and a larger simulation box (green).}

\end{figure}

In the calculations reported in Fig. \ref{VACF.hybrid}, the I-DSMC
fluid has a density (volume fraction) $\phi=0.5$ and collision frequency
prefactor $\chi=0.62$. The adiabatic sound speed is $c_{s}=\sqrt{5k_{B}T/\left(3m\right)}$
and viscosity is $\eta=\tilde{\eta}D^{-2}\sqrt{mk_{B}T}$, where we
measured $\tilde{\eta}\approx0.75$. Note that in atomistic time units
$t_{0}=D\sqrt{m/k_{B}T}$ the viscous time scale is $t_{visc}/t_{0}\approx6\phi(R_{H}/D)^{2}/(3\pi\tilde{\eta})\approx0.4(R_{H}/D)^{2}$.

As a first test case, in Fig. \ref{VACF.hybrid.small} we compare
against the particle data from Ref. \citet{SHSD_PRL}, for which the
size of the bead is $R=1.25D$, $M=7.81m$, and the simulation box
is $L=1=25D$, which corresponds to $24^{3}$ micro cells and about
$N\approx1.5\cdot10^{4}$ particles. The hybrid runs used macro cells
each composed of $4^{3}$ micro cells, which corresponds to about
$N_{0}=80$ particles per cell, and the size of the particle subdomain
fluctuated between about $3\cdot10^{3}$ and $6\cdot10^{3}$ particles
due to the change of the location of the bead relative to the continuum
grid. The particle result in Fig. \ref{VACF.hybrid.small} is the
average over 10 runs, each of length $T/t_{visc}\approx2\cdot10^{5}$,
while the hybrid results are from a single run of length $T/t_{visc}\approx7.5\cdot10^{5}$.
It is seen in the figure that both the deterministic and the fluctuating
hybrid reproduce the particle results closely, with a small but visible
difference at long times where the deterministic hybrid under-predicts
the magnitude of the tail in the VACF. We also show results from a
hybrid run with a twice larger simulation box, $L=2$, which marginally
increases the computational effort in the hybrid runs, but would increase
the length of purely particle runs by an order of magnitude. The hydrodynamic
tail becomes pronounced and closer to the theoretical prediction for
an infinite system, as expected. We have observed (not shown) that
using small continuum cells composed of only $3^{3}$ micro cells,
which corresponds to about $N_{0}=35$ particles per cell, leads to
an over-prediction of the magnitude of the VACF at short times, that
is, to an excess kinetic energy for the bead by as much as $20\%$,
depending on the exact parameters used.

As a second, more difficult test, in Fig. \ref{VACF.hybrid.huge}
we report results from particle simulations for a much larger bead,
$R=6.25D$, $M=976m$, and the simulation box is $L=2=50D$, which
corresponds to $N\approx1.2\cdot10^{5}$ particles. We have performed
a variety of hybrid runs and in Fig. \ref{VACF.hybrid.huge} we report
results from runs with macro cells each $3^{3}$ micro cells, as well
as results for a larger simulation box, $L=3$, and macro cells each
composed of $4^{3}$ micro cells. The particle results are the average
over 5 runs each of length $T/t_{visc}\approx2.5\cdot10^{3}$, while
the hybrid results are from a single run of length $T/t_{visc}\approx7.5\cdot10^{3}$.
The hybrid runs had a particle subdomain containing about $2\cdot10^{4}$
particles. We observed little impact of the size of the continuum
cell size or the size of the particle subdomain. The results in Fig.
\ref{VACF.hybrid.huge} show that the stochastic hybrid correctly
reproduces the tail in the VACF, while it slightly over-estimates
the VACF at short times. The deterministic hybrid, on the other hand,
strongly under-estimates the magnitude of $C(t)$ at both short and
long times. It is particularly striking that the deterministic hybrid
fails to reproduce the magnitude of the long time tail (and thus the
diffusion coefficient), vividly demonstrating the importance of including
fluctuations in the continuum domain.

Computing the VACF for a diffusing bead has become a standard test
for micro- and nano-scale fluid-structure coupling methods and has
been performed for a suspended bead in a wide range of particle and
(semi-)continuum compressible and incompressible fluids \citet{VACF_Alder,BrownianLB_VACF,MPCD_VACF,FluctuatingHydro_FluidOnly,BrownianSRD_Review,BrownianParticle_SIBM,BrownianParticle_IncompressibleSmoothed,SHSD_PRL}.
However, these tests often do not correctly test for all of the components
necessary to match the VACF over all relevant timescales: equipartition,
a fluctuation-dissipation relation, and hydrodynamics. Purely continuum
fluid methods allow for using a much larger time step than particle
(and thus hybrid) methods, especially if an incompressible formulation
is directly coupled to the equations of motion of the suspended bead
\citet{FluctuatingHydro_FluidOnly,BrownianParticle_SIBM,BrownianParticle_IncompressibleSmoothed}.
When fluctuations are not included in continuum methods, the VACF
is often obtained by considering the deterministic decay of the velocity
of a bead. This however assumes \emph{a priori} that proper thermodynamic
equilibrium exists with the correct fluctuation-dissipation relation,
without actually testing this explicitly. Alternatively, coupling
methods between a continuum fluid and a suspended particle often have
some arbitrary coupling parameters that are tuned to reproduce the
desired diffusion coefficient without producing a physically-consistent
VACF, especially at short times \citet{VACF_IncompressibleSmoothed}.
In particular, incompressible formulations cannot reproduce the initial
value or the decay of the VACF and should instead aim to produce an
average kinetic energy of the bead of $k_{B}T$ rather than the statistical
mechanics result of $3k_{B}T/2$ \citet{VACF_LagrangeTheory}. It
is therefore important to more carefully access the ability of other
methods in the literature to correctly reproduce the full VACF for
a truly equilibrium simulation with a large bead.

\subsection{\label{SectionPiston}Adiabatic Piston}

The problem of how thermodynamic equilibrium is established has a
long history in statistical mechanics \citet{AdiabaticPiston_Lieb}.
The \emph{adiabatic piston problem} is one of the examples used to
study the fluctuation-driven relaxation toward equilibrium \citet{AdiabaticPiston_EDMD,AdiabaticPiston_VACF,Piston_DetHydro,Piston_RelaxationModel}
that is simple enough to be amenable to theoretical analysis but also
sufficiently realistic to be relevant to important problems in nano-science
such as Brownian motors \citet{NonEqThermo_micro,BrownianMotors}.
We study the following formulation of the adiabatic piston problem.
A long quasi one-dimensional box with adiabatic walls is divided in
two halves with a thermally-insulating piston of mass $M\gg m$ that
can move along the length of the box without friction. Each of the
two halves of the box is filled with a fluid that is, initially, at
a different temperature $T$ and density $\rho$, here assumed to
follow the ideal equation of state $P=\rho k_{B}T/m$. If the macroscopic
pressures in the two halves are different, $\rho_{L}T_{L}\neq\rho_{R}T_{R}$,
then the pressure difference will push the piston to perform macroscopic
oscillations with a period that can be estimated by assuming that
each half undergoes an adiabatic transformation ($PV^{\gamma}=\mbox{const.}$).
These oscillations are damped by viscous friction and lead to the
piston essentially coming to rest in a state of \emph{mechanical equilibrium},
$\rho_{L}T_{L}=\rho_{R}T_{R}$. This stage of the relaxation from
non-equilibrium to mechanical equilibrium has been shown to be well-described
by deterministic hydrodynamics \citet{Piston_DetHydro}.

The state of mechanical equilibrium is however not a state of true
thermodynamic equilibrium, which also requires equality of the temperatures
on the two sides of the piston. Reaching full equilibrium requires
heat transfer through the piston, but the piston is adiabatic and
does not conduct heat. In classical deterministic hydrodynamics the
piston would just stand still and never reach full equilibrium. It
has been realized long ago that heat is slowly transferred through
the mechanical asymmetric fluctuations of the piston due to its thermal
motion, until the temperatures on both sides of the piston equilibrate
and the fluctuations become symmetric. This equilibration takes place
through a single degree of freedom (the piston position) coupling
the two large reservoirs, and it would be astronomically slow in a
laboratory setting. While various Langevin or kinetic theories have
been developed for the effective heat conduction of the adiabatic
piston (see Refs. \citet{AdiabaticPiston_EDMD,AdiabaticPiston_VACF,Piston_DetHydro,Piston_RelaxationModel}
and references therein), there is no complete theoretical understanding
of the effective heat conductivity, especially in dense fluids. Molecular
dynamic simulations have been performed in the past \citet{AdiabaticPiston_EDMD,AdiabaticPiston_VACF,Piston_DetHydro}
using hard-disk fluids, but the very long runs required to reach thermodynamic
equilibrium for massive pistons have limited the size of the systems
that could be studied. Furthermore, there have been no studies applying
fluctuating hydrodynamics to this problem.

\begin{figure}[tbph]
\begin{centering}
\includegraphics[width=0.95\textwidth]{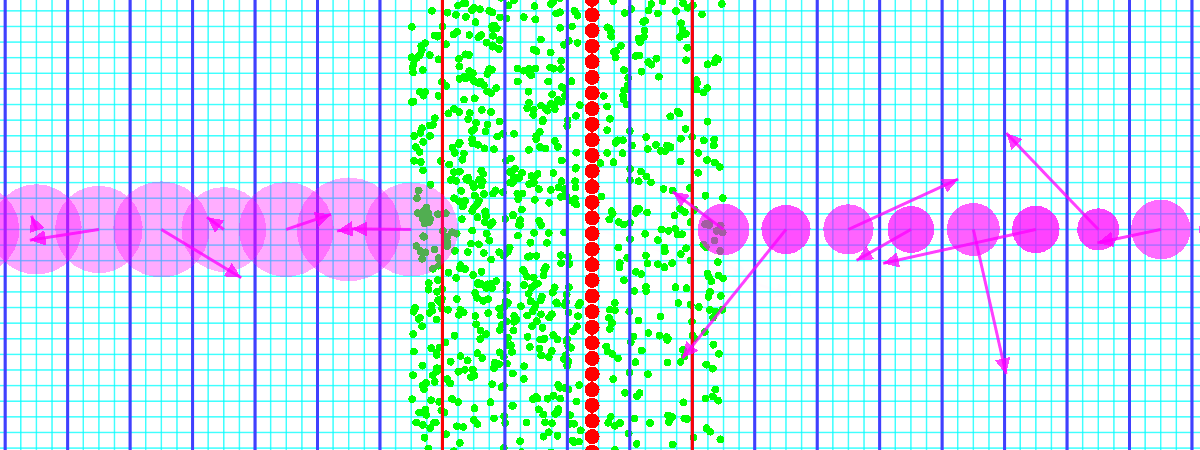}
\par\end{centering}

\caption{\label{fig:PistonSetup}An illustration of the computational setup
used for the adiabatic piston computations. Only the central portion
of the box of aspect ratio $6\times1$ is shown. Left of the piston
the gas is cold and dense; to the right it is hot and dilute. The
piston beads (red disks) separate the box into two halves, and are
surrounded on each side by a fluid of I-DSMC particles (smaller green
disks), which is twice denser but also twice cooler in the left half
than in the right half. The microscopic grid is shown with thinner
light blue lines and the hydrodynamic grid is shown with thicker dark
blue lines. The interface between the particle and continuum regions
is highlighted with a thick red line. A snapshot of the values of
the hydrodynamic variables in each continuum cell is shown using a
large purple disk whose size is proportional to the density and its
opaqueness is proportional to the temperature, and an arrow for the
fluctuating velocities.}

\end{figure}

Here we apply our hybrid method to the adiabatic piston problem in
two dimensions, using a non-linear two-dimensional implementation
of the Runge-Kutta integrator described in Ref. \citet{LLNS_S_k}
as the continuum solver. The choice of two dimensions is for purely
computational reasons. Firstly, the number of particles required to
fill a box of sufficient size is much smaller thus allowing for long
particle simulations. Secondly, in order to implement the piston in
our particle scheme we reused the same mixed event-driven/time-driven
handling \citet{DSMC_AED} as we used for the VACF computations in
Section \ref{SectionVACF}. Namely, we made a piston out of $N_{b}$
small impermeable beads, connected together to form a barrier between
the two box halves, as illustrated in Fig. \ref{fig:PistonSetup}.
In two dimensions, by ensuring that two piston beads never separate
by more than a given distance we can ensure that two I-DSMC particles
on opposite sides of the piston cannot possibly collide and thus the
piston will be insulating. We have studied two different types of
pistons, a \emph{flexible} piston where the the beads are tethered
together to form a chain \citet{DSMC_AED} that is stretched but where
each individual bead can still move independently of the others, and
a \emph{rigid} piston that is obtained with a slight modification
of the event loop to move all of the piston beads in unison. While
at the macroscopic level the exact shape of the piston should not
make a big difference, we have found that increasing the number of
degrees of freedom of the piston from one to $N_{b}$ makes a significant
difference in the thermal conductivity of the piston, and therefore,
we will focus here on rigid pistons as in the traditional formulation.
We use specular collisions of the fluid particles with the piston
beads, although qualitatively identical (but not quantitatively identical)
results are obtained using bounce-back collisions as well.

The hybrid method setup for the adiabatic piston is illustrated in
Fig. \ref{fig:PistonSetup}. We use a two-dimensional Maxwell I-DSMC
particle fluid ($\phi=1$, $\chi=1$) with collision diameter $D=0.1$
(hard-sphere diameter $D_{s}=D/2$) and a piston composed of $N_{b}=40$
beads of diameter $D_{b}=0.0955$. The particle subdomain is limited
to a few continuum cells around the piston, which we keep at about
two or more continuum cells on each side of the piston, so that the
unreasonable hydrodynamic values in the cells that overlap the piston
do not affect the continuum solver appreciably. Periodic boundary
conditions are applied along the $y$ dimension (parallel to the piston)
with the width of the domain $L_{y}=4$ being $40$ microscopic cells,
while adiabatic walls were placed at the ends of the box whose total
length $L_{x}=24$ was $240$ microscopic cells. We have studied various
sizes for the macroscopic cells, and report results for a quasi one-dimensional
continuum grid in which each macro cell contains $4\times40$ micro
cells, corresponding to about $200$ particles per continuum cell.
We also present results for a two-dimensional continuum grid where
each macro cell contains $8\times10$ micro cells.

\begin{figure}[tbph]
\begin{centering}
\includegraphics[width=0.75\textwidth]{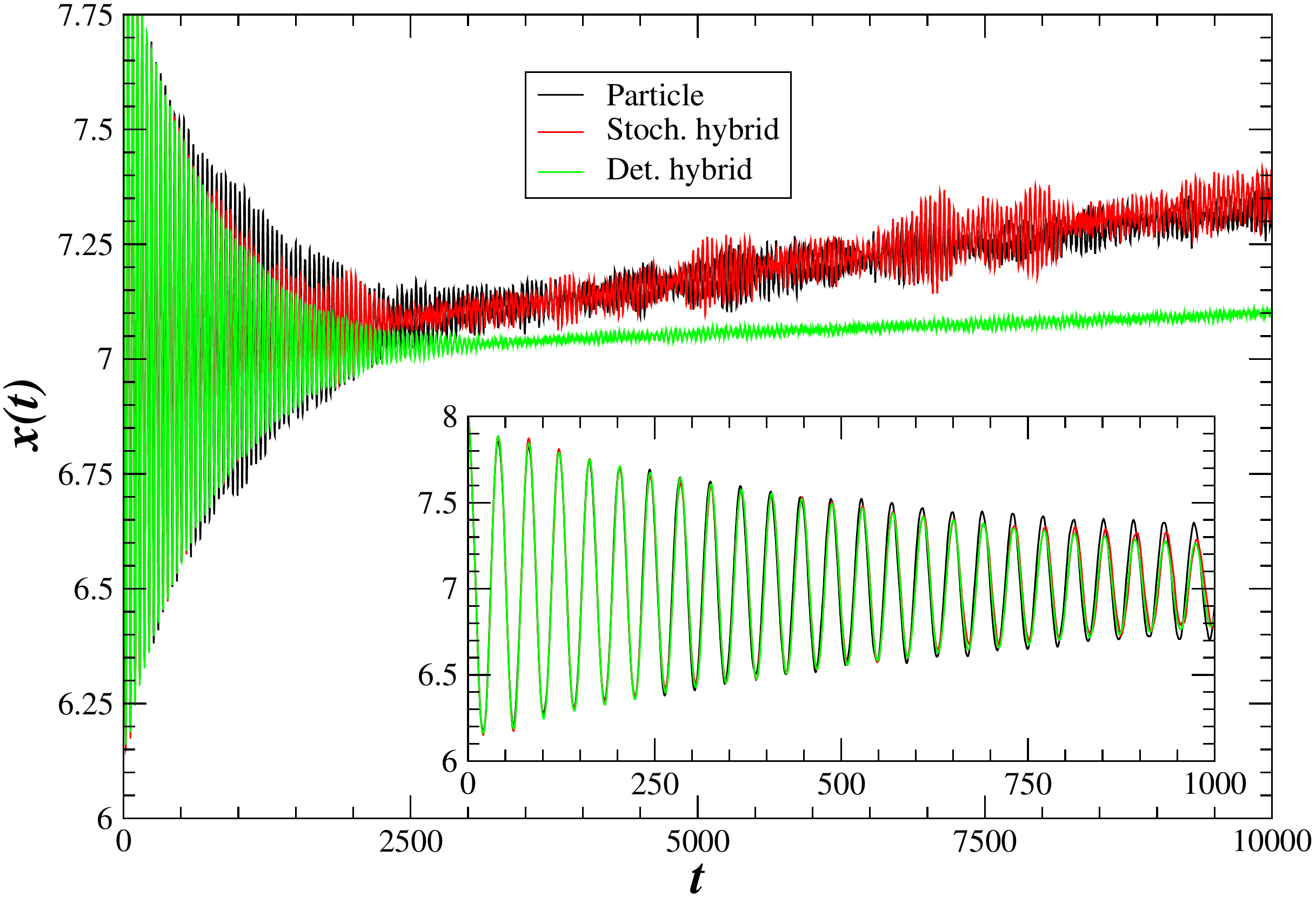}
\par\end{centering}

\caption{\label{fig:MassivePiston}Relaxation of a massive rigid piston ($M/m=4000$)
from a position $x=8$ that is not in mechanical equilibrium. Through
rapid damped oscillations mechanical equilibrium is established at
position $x\approx7$, after which a slow relaxation to true equilibrium
is seen. The stochastic hybrid is able to match the particle data
very well, to within the expected statistical difference from realization
to realization. The deterministic hybrid, on the other hand, clearly
under-predicts the rate of relaxation. The inset highlights the initial
oscillations and shows that in the regime where fluctuations do not
matter the deterministic and stochastic hybrids do not differ appreciably.}

\end{figure}

We have performed hybrid runs with both the deterministic and stochastic
hybrids. In Fig. \ref{fig:MassivePiston} we show the position of
a massive piston of mass $M=4000m$ that started at a position $x=8$
that is not in mechanical equilibrium and thus performs rapid damped
oscillations until it reaches a state of mechanical equilibrium at
$x\approx7$, from which it slowly relaxes toward true equilibrium.
The results in the figure show that the stochastic hybrid reproduces
the correct relaxation toward equilibrium while the deterministic
hybrid severely under-predicted the rate of equilibration (effective
heat conductivity), even though the initial mechanical stage of the
relaxation is correctly captured by both hybrids, as expected. We
have observed that the deterministic hybrid fails to give the correct
answer whenever a rigid massive piston is used, $M>250m$. For flexible
pistons, we find that even for a large bead mass $M_{b}$(overall
piston mass $M=N_{b}M_{b}$) both the deterministic and stochastic
hybrids reproduce the purely particle results for the slow relaxation
toward equilibrium.

\begin{figure}[tbph]
\begin{centering}
\includegraphics[width=0.7\textwidth]{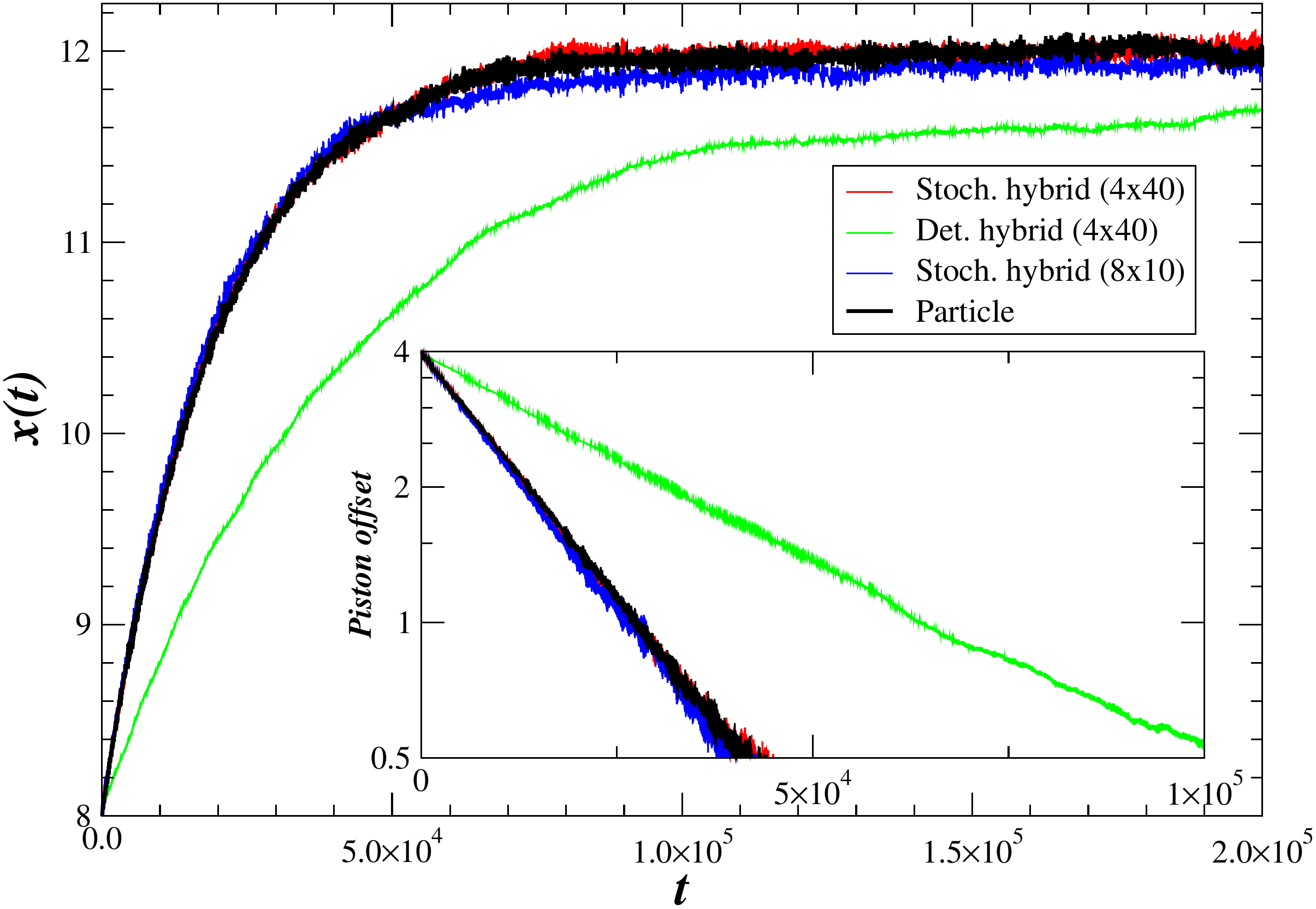}
\par\end{centering}

\caption{\label{fig:PistonRelaxation}Relaxation of a rigid piston of mass
$M/m=1000$ from an initial state of mechanical equilibrium ($x=8$)
to a state of thermodynamic equilibrium ($x=12$). The inset emphasizes
the initial exponential decay on a semi-log scale. The hybrid runs
used a particle subdomain of width $w_{P}=2$ on each side of the
piston and continuum cells that were composed of either $4\times40$
or $8\times10$ microscopic cells. For the deterministic hybrid the
macro cell size makes little difference so we only show the $4\times40$
case.}

\end{figure}

A more detailed comparison of the particle and hybrid results for
a piston of mass $M=1000m$ that is initially in mechanical equilibrium
at position $x=8=L_{x}/3$ is shown in Fig. \ref{fig:PistonRelaxation}.
The initial conditions were $k_{B}T_{L}=2/3$, $\rho_{L}=2/3$ and
$k_{B}T_{R}=4/3$, $\rho_{L}=1/3$, so that there is an equal mass
on each side of the piston. At the true equilibrium state the piston
remains close to the middle of the box, $x_{eq}=L/2=12$, with equal
density on each side. The results shown are averages over 10 samples,
but it should be emphasized that each run exhibits thermal oscillations
of the piston%
\footnote{At thermodynamic equilibrium with a common temperature $T$, the frequency
of the thermally-driven oscillations can be estimated using a quasi-adiabatic
harmonic approximation to be $\omega^{2}\approx Nk_{B}T/\left[Mx(L_{x}-x)\right]$
(see also an alternative derivation in Ref. \citet{AdiabaticPiston_VACF})
and the amplitude of the oscillations can be estimated to be on the
order of $\D{x}^{2}\approx x(L_{x}-x)/N$, where $N$ is the number
of fluid particles per chamber.%
} position that are diminished by direct averaging because they have
different (random) phases. Figure \ref{fig:PistonRelaxation} shows
that the stochastic hybrid is able to correctly reproduce the rate
of exponential relaxation of the piston toward equilibrium with many
fewer particles than the purely particle runs, while the deterministic
hybrid fails. We have observed a slight dependence on the exact details
of the hybrid calculations such as cell size or the width of the particle
subdomain; however, in general, the stochastic hybrid has shown to
be remarkably robust and successful. At the same time, the importance
of including thermal fluctuations in the continuum subdomain is revealed
as for the VACF computations in Section \ref{SectionVACF}.

\begin{figure}[tbph]
\begin{centering}
\includegraphics[width=0.7\textwidth]{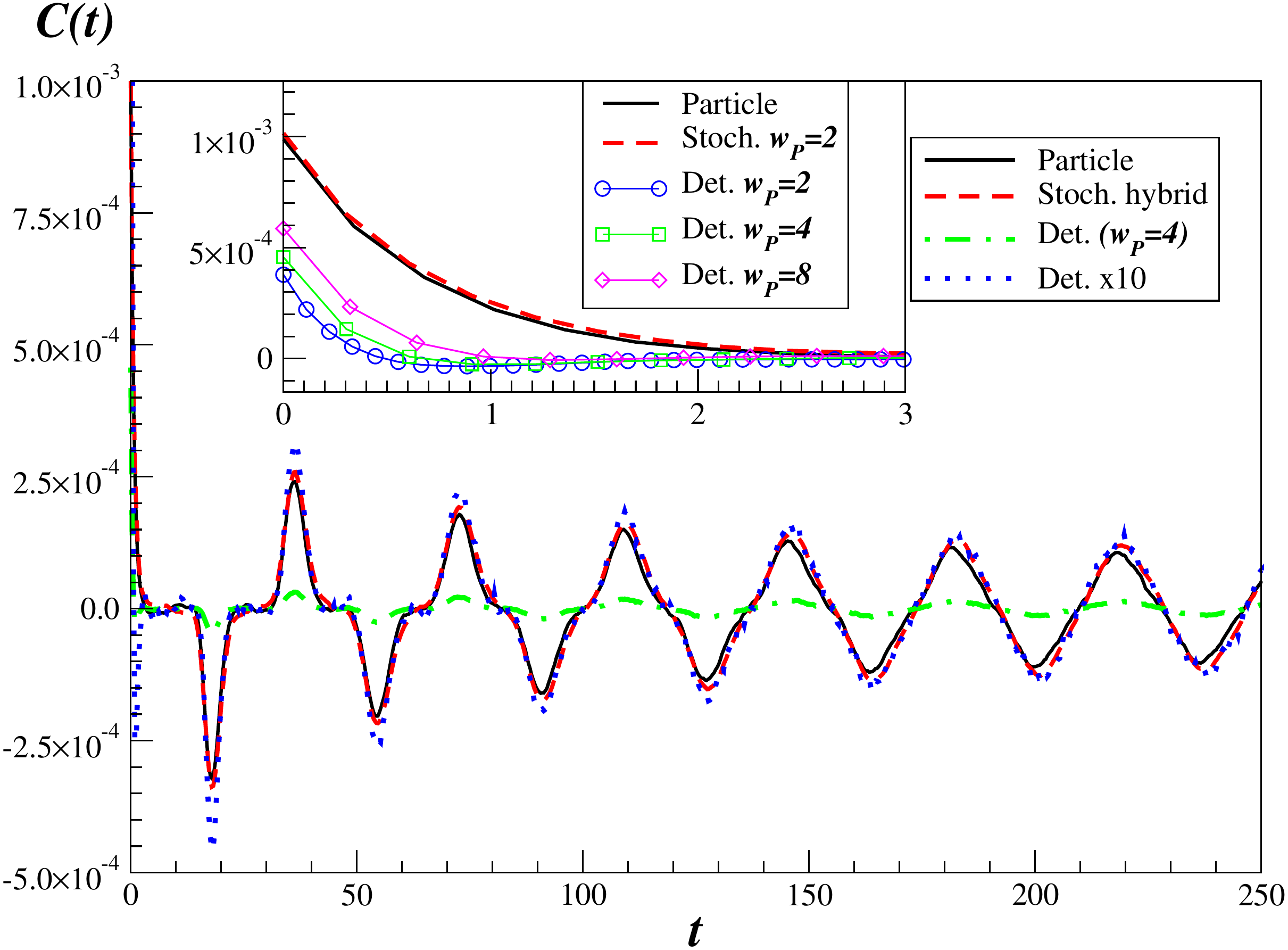}
\par\end{centering}

\caption{\label{fig:PistonVACF}The velocity autocorrelation function for a
rigid piston of mas $M/m=1000$ at thermal equilibrium at $k_{B}T=1$
(thus, the expected initial value is $C(0)=10^{-3}$), and macro cells
of size $4\times40$ micro cells. The deterministic hybrid gives much
smaller effective temperature $T_{{\rm eff}}$ of the piston and a
negative dip in $C(t)$ at short times, but when magnified by an order
of magnitude it reveals the correct shape at longer times. The inset
focuses on the initial decay of the VACF and shows that even increasing
the width of the particle region to $w_{P}=8$ (the whole box has
a length of $60$ continuum cells) does not help the deterministic
hybrid much whereas the stochastic hybrid gives the correct initial
decay.}

\end{figure}

The relation to the equilibrium VACF computations in Section \ref{SectionVACF}
is emphasized by computing the VACF $C(t)=\left\langle v_{x}(t)v_{x}(0)\right\rangle $
for the piston in its state of true equilibrium%
\footnote{In Ref. Ref. \citet{AdiabaticPiston_VACF} the autocorrelation function
for the piston bead position is used to extract the rate of exponential
rate toward true equilibrium, however, a direct non-equilibrium calculation
as we perform in Fig. \ref{fig:PistonRelaxation} is more efficient
and illustrative for our purpose.%
}, $k_{B}T=1$. The VACF is rather complex due to the interplay of
short-time kinetic effects, dissipation, and sound reflections from
the walls, however, our focus here is to simply compare against the
purely particle simulations and not to understand all the features
in the VACF. The results, shown in Fig. \ref{fig:PistonVACF}, reveal
that the piston does not equilibrate at the correct effective temperature
$T_{{\rm eff}}=MC(0)/k_{B}$ in the deterministic hybrid calculations.
Notably, just as we found for the massive bead in Section \ref{SectionVACF},
the piston has a kinetic energy that is markedly lower (half) than
the correct value $C(0)=k_{B}T/M=10^{-3}$ when fluctuations are not
consistently included in the continuum region. Since the quasi-equilibrium
temperature of the piston%
\footnote{It is predicted that the piston equilibrates at a temperature that
is approximately the geometric mean of the left and right temperatures
\citet{AdiabaticPiston_Langevin}.%
} plays a crucial role in all of the kinetic theories \citet{AdiabaticPiston_EDMD,AdiabaticPiston_VACF,Piston_RelaxationModel}
for the effective heat transfer, it is not surprising that the deterministic
hybrid gives the wrong answer. What is even more striking is that
increasing the width of the particle subdomain $w_{P}$ on each side
of the piston, as measured in units of continuum cells (Fig. \ref{fig:PistonSetup}
shows the case $w_{P}=2$), barely improves the accuracy of the VACF
at short times, showing that the whole spectrum of fluctuations affects
the initial rate of decay of the piston. At the same time, the deterministic
hybrid does give the correct shape of the VACF at longer times, as
revealed by magnifying the VACF for $w_{P}=4$ by an arbitrary factor
of $10$ to bring it in close agreement with the particle result at
longer times%
\footnote{The magnification factor required to bring the long-time VACF for
the deterministic hybrid in agreement with the correct result decreases
with increasing $w_{P}$, for example, it is $\sim20$ for $w_{P}=2$
and $\sim4$ for $w_{P}=8$.%
}. This is also not unexpected since the VACF at longer times is dominated
by mechanical vibrations and dissipation, and is analogous to what
we find for the dynamic structure factor when it is calculated by
a deterministic hybrid scheme.

\section{Conclusions}

We have described a hybrid particle-continuum algorithm for simulating
complex flows and applied it to several non-trivial problems. The
algorithm couples an ideal stochastic particle fluid algorithm with
a fluctuating hydrodynamic continuum solver using a direct dynamic
coupling where the continuum solver supplies Dirichlet-like (state)
boundary conditions for the particle region, while the particle region
supplies von Neumann-like (flux) boundary conditions to the continuum
solver. The continuum solver computes a provisional solution over
the whole domain that then gets replaced with the particle data in
the particle region and also gets corrected (refluxed) in the cells
bordering the particle subdomain. We described the components necessary
to extend previous variants of this rather general coupling methodology
\citet{AMAR_DSMC,AMAR_DSMC_SAMRAI,FluctuatingHydro_AMAR} to use a
recently-developed variant of the DSMC particle method suitable for
dense fluids \citet{SHSD_PRL,SHSD}, as well as an explicit conservative
compressible fluid solver that accurately accounts for thermal fluctuations
in the Navier-Stokes equations \citet{LLNS_S_k}. By turning the fluctuating
fluxes off in the continuum solver we can trivially transform our
stochastic hybrid method to a deterministic hybrid method closer to
commonly-used hybrid schemes.

In Section \ref{sec:Results} we applied our stochastic hybrid method
to several challenging problems and demonstrated that it could obtain
the correct answer with significantly fewer particles and thus significantly
less computational effort than a purely particle simulation. We used
purely particle runs as a {}``gold'' standard against which we judge
the accuracy of the hybrid method, consistent with using the hybrid
method for situations in which a purely continuum description is unable
to capture the full physics and a particle method is necessary in
some portion of the physical domain. For example, particle methods
such as DSMC are essential to correctly resolve the flow at a high
Mach number shock \citet{FluctuatingHydro_AMAR}, or in kinetic flow
regions such as the wake behind a fast-moving body \citet{AMAR_DSMC}
or small channels in a micro-electromechanical device \citet{GasFlows_Nicolas}.

In Section \ref{SectionMismatch} we showed that there is a small
mismatch between the density and temperature in the particle and continuum
regions, caused by using instantaneous fluctuating values of the hydrodynamic
variables when generating velocities for the reservoir particles,
and that the effect is of order $N_{0}^{-1}$, where $N_{0}$ is the
average number of particles in one cell. In Appendix \ref{AppendixFluxMismatch}
we demonstrated that this is not an artifact of the method but rather
of an inherent difficulty in stochastic methods where instantaneous
values are used to estimate means. In section \ref{Section_S_kw}
we calculated the dynamic structure factor using the hybrid method
for a quasi two-dimensional situation and a large wavevector $\V{k}$
that is obliquely incident to the particle-continuum interface, and
confirmed that spontaneous sound and entropic fluctuations are transmitted
correctly through the interface. We also calculated the dynamic structure
factor for a finite system bounded by two adiabatic walls and observed
excellent agreement with theoretical calculations given in Appendix
\ref{Appendix_S_kw_walls}.

In Section \ref{SectionVACF} we studied the diffusive motion of a
large spherical neutrally-buoyant bead suspended in a fluid of I-DSMC
particles in three dimensions by placing a mobile particle subdomain
around the suspended bead. We computed the velocity autocorrelation
function (VACF) for two bead sizes using the stochastic and deterministic
hybrids as well as purely particle simulations. We found that the
stochastic hybrid correctly reproduces the VACF over all time scales,
while the deterministic hybrid under-estimates both the kinetic energy
of the bead and the magnitude of the tail of the VACF for sufficiently
large beads. Finally, in Section \ref{SectionPiston} we applied the
hybrid scheme to a non-equilibrium quasi one-dimensional version of
the \emph{adiabatic piston} problem \citet{AdiabaticPiston_Lieb,AdiabaticPiston_EDMD},
a classic example of the importance of fluctuations in establishing
global thermal equilibrium. The two-dimensional particle region was
placed around the piston and an event-driven algorithm was used to
handle the interaction of the fluid with the piston. We again found
that the stochastic hybrid was able to reproduce the purely particle
results correctly, while the deterministic calculations under-estimated
the relaxation substantially for sufficiently massive rigid pistons.

Our results for both the VACF and the adiabatic piston clearly demonstrated
that a large massive suspended body cannot equilibrate at the correct
Boltzmann distribution unless thermal fluctuations are consistently
included in the full domain, including the continuum region in hybrid
methods, even if a large particle subdomain is used. This points to
an increased importance of the long-wavelength, and thus slowly-decaying,
hydrodynamic fluctuations. Massive and large suspended bodies have
longer relaxation times and thus it is not surprising that slower-decaying
fluctuations play a more prominent role for them than for smaller
suspended particles. However, a better theoretical understanding of
these observations is necessary in order to establish the importance
of hydrodynamic fluctuations in general. At the same time, our results
make it clear that fluctuations should be included in the continuum
region in hybrid methods, consistent with the particle dynamics, rather
than treating the fluctuations as {}``noise'' from which the continuum
solver ought to be shielded.

Fluctuating hydrodynamics has successfully accounted for thermal fluctuations
in a variety of problems. At the same time, however, the nonlinear
stochastic partial differential Landau-Lifshitz Navier-Stokes equations
are mathematically ill-defined and require a cutoff length and/or
time scale to be interpreted in a reasonable sense. The linearized
equations can be given a well-defined interpretation, however, they
are physically unsatisfactory in that they require a base state to
linearize around which may itself be time-dependent and unknown or
even be affected by fluctuations, as in the adiabatic piston or diffusing
shock problems \citet{AMAR_Burgers,FluctuatingHydro_Garcia}. We have
numerically observed that using small continuum cells leads to worse
results even if the linearized LLNS equations are used, thus formally
avoiding the difficulties with the increased relative magnitude of
the fluctuations. This is consistent with the expectation that a continuum
description is only applicable on length scales and time scales sufficiently
larger than the molecular size and molecular collision time. In our
experience using more than $75$ particles per cell leads to a good
match between the hybrid and particle runs; however, a better theoretical
understanding of the proper inclusion of fluctuations in hydrodynamics
is a necessary future development.

Our implementation is at present serial and our runs are therefore
limited by the CPU-intensive collision procedure in the I-DSMC particle
algorithm. Some of the examples we presented utilized the mixed event-
and time-driven particle algorithm developed in Ref. \citet{DSMC_AED}.
The event-driven component of this algorithm is notoriously difficult
to parallelize. However, a purely time-driven particle algorithm can
easily be parallelized, as can the purely continuum solver. We plan
to implement a parallel hybrid scheme in the future in order to enable
the study of realistic system sizes. At the same time, however, reaching
long time scales will necessarily require time steps beyond the small
ones required by particle methods and explicit continuum schemes.

\begin{appendix}

\section{\label{AppendixFluxMismatch}Finite-size mismatch between particle
and continuum descriptions}

Consider a simple particle/continuum hybrid consisting of a one-dimensional
system with an ideal particle fluid on one side and a fluctuating
continuum solver on the other. At equilibrium the mean density and
temperature are $\rho_{0}$ and $T_{0}$, respectively; the mean fluid
velocity is taken as zero. The problem we want to consider is whether
the one-sided fluxes of mass, momentum, and energy due to the reservoir
particles that enter the continuum subdomain are correct, assuming
that the means and variances of the hydrodynamic variables in the
reservoir cells are correct.

In the continuum calculation instantaneous density and temperature,
$\rho$ and $T$, are computed and used to generate particles for
injection. The one-sided fluxes of mass, momentum, and energy are
known from kinetic theory to be \begin{eqnarray*}
F_{\rho}(\rho,T) & = & \sqrt{\frac{k}{2\pi m}}\,\rho T^{1/2}\\
F_{j}(\rho,T) & = & \frac{k}{2m}\,\rho T\\
F_{E}(\rho,T) & = & \sqrt{\frac{4k^{3}}{m^{3}\pi}}\,\rho T^{3/2}.\end{eqnarray*}
The correct mean equilibrium mass flux is $F_{\rho}(\rho_{0},T_{0})$;
however, the mean value of the particle flux does not equal this correct
value since \[
\langle\rho T^{1/2}\rangle=\langle\rho\rangle\langle T^{1/2}\rangle\neq\langle\rho\rangle\langle T\rangle^{1/2};\]
note that density and temperature are instantaneously uncorrelated.
Similarly, the mean energy flux is also incorrect; interestingly the
momentum flux is correct since it has a linear dependence on temperature.

To find the errors in the fluxes, we write $T=T_{0}+\delta T$ and
for a given power exponent $a$ we have \[
\langle T^{a}\rangle=\langle T_{0}^{a}\rangle\langle(1+\delta T/T_{0})^{a}\rangle\approx\langle T_{0}^{a}\rangle\left[1+{\textstyle \frac{1}{2}}a(a-1)\frac{\langle\delta T^{2}\rangle}{T_{0}^{2}}\right]=\langle T_{0}^{a}\rangle\left[1+\frac{a(a-1)(\gamma-1)}{2N_{0}}\right],\]
where $N_{0}=\rho_{0}V_{c}/m$ is the number of particles in a continuum
cell (of volume $V_{c}$). From this we have that, for a monatomic
gas ($\gamma=5/3$), \begin{eqnarray*}
\langle F_{\rho}(\rho,T)\rangle & = & F_{\rho}(\rho_{0},T_{0})\left(1-\frac{1}{12N_{0}}\right)\\
\langle F_{J}(\rho,T)\rangle & = & F_{J}(\rho_{0},T_{0})\\
\langle F_{E}(\rho,T)\rangle & = & F_{E}(\rho_{0},T_{0})\left(1+\frac{1}{4N_{0}}\right).\end{eqnarray*}
This result shows that there is no way for all three mean fluxes to
be correct when the particles are generated from the Maxwell-Boltzmann
distribution using instantaneous, fluctuating values of density and
temperature. The mismatch is of the order $N_{0}^{-1}$, where $N_{0}$
is the number of particles per macro cell, and therefore the mismatch
gets worse as the macroscopic cells become smaller and the (relative)
fluctuations become larger. As our numerical results show, because
of this mismatch between the particle and hydrodynamic descriptions
it will be impossible for the particle and continuum regions to reach
a common equilibrium state.

\section{\label{Appendix_S_kw_walls}Dynamic Structure Factor With Adiabatic
Walls}

Dynamic structure factors are easily calculated in the bulk by using
the spatio-temporal Fourier transform of the linearized LLNS equations
\citet{FluctHydroNonEq_Book}. For finite domains, such as slab geometries,
there are results in the literature but they are often restricted
to some simplified models or complex non-equilibrium situations \citet{DSMC_Fluctuations_Shear,DynamicStructureFactor_walls}.
We therefore derive here the equilibrium dynamic structure factors
for a fluid in-between two adiabatic hard walls by solving the LLNS
equations with the appropriate boundary conditions.

Boundary conditions change the Hilbert space in which a solution is
to be sought and the corresponding basis functions (eigenfunctions
of the generator with the specified BCs). For the LLNS equations with
adiabatic boundaries (i.e., slip insulating walls) along the $x$
axes, the appropriate basis functions are $\cos(kx)$ for density
and temperature, where $k=p\pi/L$ is the wavevector enumerated by
the wave index $p\in\Set{Z}^{+}$, and $\sin(kx)$ for the velocities
\citet{DSMC_Fluctuations_Shear,DynamicStructureFactor_walls}, as
compared to the ones for {}``bulk'' conditions (periodic boundaries),
$\exp(2q\pi x/L)$, with $k=2q\pi/L$ and wave index $q\in\Set{Z}$
\citet{LLNS_S_k,FluctHydroNonEq_Book}. For thermal walls (constant-temperature
stick boundaries) there does not appear to be a simple basis.

White noise has a trivial expansion in either the sine or cosine basis
sets, namely, all of the coefficients are i.i.d. Gaussian random variables
with mean zero and variance $2$. The generator of the Navier-Stokes
equations separates wavevectors/frequencies into the the same $(k,\omega)$
equations as for {}``bulk'' (periodic BCs), and therefore the dynamic
structure factor, if expressed in the given basis set, has the same
familiar form \citet{DSMC_Fluctuations_Shear,DynamicStructureFactor_walls}.
In particular,\[
\rho(x,t)=\rho_{0}+\sum_{p=1}^{\infty}\rho_{p}(t)\cos(p\pi x/L)=\rho_{0}+\sum_{p=1}^{\infty}\int_{-\infty}^{\infty}d\omega e^{-i\omega t}\rho_{p,\omega}\cos(p\pi x/L),\]
where the different $p$'s and $\omega$'s are uncorrelated\[
\left\langle \rho_{p,\omega}\rho_{p^{\prime},\omega^{\prime}}^{\star}\right\rangle =2\tilde{S}_{p,\omega}\delta_{p,p^{\prime}}\delta\left(\omega-\omega^{\prime}\right),\]
where $\tilde{S}_{p,\omega}=\tilde{S}_{k=p\pi/L,\omega}$ denotes
the usual bulk dynamic structure factor.

Now, we need to convert this result to the more usual Fourier basis,
$\exp(2q\pi x/L)$, since this is how dynamic structure factors are
defined. From the Fourier inversion formula, and the orthogonality
of the cosine basis functions, we have\[
\rho_{q}=\frac{1}{L}\int_{0}^{L}dx\rho(x,t)e^{-2q\pi x/L}=\frac{1}{2}\rho_{p=2q}-\frac{i}{L}\sum_{p=1}^{\infty}\rho_{p}\int_{0}^{L}dx\cos(p\pi x/L)\sin(2q\pi x/L).\]
By performing the integration explicitly we get\[
\rho_{q}=\frac{1}{2}\rho_{p=2q}+\frac{4iq}{\pi}\sum_{p\;{\rm odd}}\frac{\rho_{p}}{p^{2}-4q^{2}},\]
giving the dynamic structure factor\begin{equation}
S_{q,\omega}^{(\rho)}\equiv S_{k=2\pi q/L,\omega}^{(\rho)}=\left\langle \rho_{q,\omega}\rho_{q,\omega}^{\star}\right\rangle =\frac{1}{2}\tilde{S}_{k,\omega}^{(\rho)}+\frac{32q^{2}}{\pi^{2}}\sum_{p\;{\rm odd}}\frac{\tilde{S}_{p,\omega}^{(\rho)}}{\left(p^{2}-4q^{2}\right)^{2}}.\label{S_kw_cosine}\end{equation}
Each of the terms under the sum gives an additional peak at $\omega=ck=pc\pi/L$,
where $p$ is odd, which arises physically because of the standing
waves that appear due to reflections of the sound waves from the walls.
Only the first few terms need to be kept to get most of the power
(the total integral over $\omega$ is one), and it can be shown (by
simple numerical comparison or explicit summation) that the above
is equivalent to the more opaque Eq. (12) in Ref. \citet{DynamicStructureFactor_walls}
(set $\gamma=0$ for no shear).

Equations identical to (\ref{S_kw_cosine}) hold for temperature and
the velocity components parallel to the wall. For the perpendicular
velocity ($v_{x}$) a sine basis is more appropriate, and a similar
calculation gives the dynamic structure factor for the finite system
in terms of the bulk one,\[
S_{k=2\pi q/L,\omega}^{(v_{\perp})}=\frac{1}{2}\tilde{S}_{k,\omega}^{(v_{\perp})}+\frac{8}{\pi^{2}}\sum_{p\;{\rm odd}}\frac{p^{2}\tilde{S}_{p,\omega}^{(v_{\perp})}}{\left(p^{2}-4q^{2}\right)^{2}}.\]

\section{\label{AppendixWallsLLNS}Handling of Adiabatic and Thermal Walls
in the Continuum Solver}

Solving the LLNS equations with non-periodic boundaries requires some
special handling of the stochastic fluxes at the boundaries, which
are assumed to coincide with faces of the continuum grid. As discussed
in Refs. \citet{AMR_ReactionDiffusion_Atzberger,LLNS_S_k}, the numerical
discretization of the Laplacian operator $\M{L}$, the divergence
operator $\M{D}$, and the gradient operator $\M{G}$ should satisfy
a \emph{discrete} fluctuation-dissipation balance condition $\M{L}=\M{D}\M{C}\M{G}=-\M{D}\M{C}\M{D}^{*}$,
where $\M{C}$ is a dimensionless covariance matrix for the stochastic
fluxes that are generated using a random number generator on each
face of the grid. For one-dimension with periodic boundaries it is
well known that the standard face-to-cell divergence, cell-to-face
gradient, and three-point Laplacian second-order stencils satisfy
$\M{L}=\M{D}\M{G}$ and thus $\M{C}=\M{I}$ (the identity matrix)
works and it is in fact trivial to implement algorithmically \citet{AMR_ReactionDiffusion_Atzberger,LLNS_S_k}.
When boundaries are present, the stencils near the boundaries are
modified to take into account the boundary conditions.

Algorithmically, \emph{ghost cells} extending beyond the boundaries
are used to implement modified finite-difference stencils near the
boundaries. The numerical scheme continues to use standard divergence
$\M{D}$ (face-to-cell) and gradient (cell-to-face) $\M{G}=-\M{D}^{\star}$
stencils but implements a modified Laplacian operator due to special
handling of the ghost cells. If a Dirichlet condition is imposed on
a given variable (e.g., a fixed wall temperature), then the ghost
cell value is obtained by a linear extrapolation of the value in the
neighboring interior cell (inverse reflection). If a von Neumann condition
is imposed, on the other hand, then the ghost cell value is set equal
to the value in the neighboring interior cell (reflection). This gives
discrete operators that can be represented by the following banded
matrices near the left corner (first cell) of a one-dimensional domain\[
\M{D}=\left[\begin{array}{cccc}
-1 & 1 & 0 & \cdots\\
0 & -1 & 1 & \ldots\\
0 & 0 & -1 & \cdots\\
\vdots & \vdots & \vdots & \ddots\end{array}\right],\quad\M{L}=\left[\begin{array}{cccc}
-(2-\alpha) & 1 & 0 & \cdots\\
1 & -2 & 1 & \ldots\\
0 & 1 & -2 & \cdots\\
\vdots & \vdots & \vdots & \ddots\end{array}\right],\]
where $\alpha=-1$ for a Dirichlet condition and $\alpha=1$ for a
von Neumann condition. It is easy to verify that $\M{L}=-\M{D}\M{C}\M{D}^{*}$
is satisfied with the following diagonal scaling matrix\[
\M{C}=\left[\begin{array}{cccc}
\beta & 0 & 0 & \cdots\\
0 & 1 & 0 & \ldots\\
0 & 0 & 1 & \cdots\\
\vdots & \vdots & \vdots & \ddots\end{array}\right],\]
where $\beta=1-\alpha.$

This direct computation shows that in order to satisfy the discrete
fluctuation-dissipation balance condition the diagonal element of
$\M{C}$ corresponding to the cell face that touches the boundary
ought to be set to $2$ for a Dirichlet and to $0$ for a von Neumann
condition. This means that the corresponding component of the stochastic
flux needs to be generated using a random normal variate of variance
$2$ for Dirichlet, and set to zero for a von Neumann condition.

Finally, for density, the ghost cell values are generated so that
the pressure in the ghost cells is equal to the pressure in the neighboring
interior cell, which ensures that there is no unphysical pressure
gradient in the momentum equation across the interface. There is also
no stochastic mass flux through faces on the boundary independent
of the type of boundary condition at the wall.

\end{appendix}


\begin{thebibliography}{10}

\bibitem{ParticleMesoscaleHydrodynamics}
H.~Noguchi, N.~Kikuchi, and G.~Gompper.
\newblock {Particle-based mesoscale hydrodynamic techniques}.
\newblock {\em Europhysics Letters}, 78:10005, 2007.

\bibitem{TripleScale_Rafael}
R.~Delgado-Buscalioni, K.~Kremer, and M.~Praprotnik.
\newblock {Concurrent triple-scale simulation of molecular liquids}.
\newblock {\em J. Chem. Phys.}, 128:114110, 2008.

\bibitem{MultiscaleMicrofluidics_Review}
G.~Hu and D.~Li.
\newblock {Multiscale phenomena in microfluidics and nanofluidics}.
\newblock {\em Chemical Engineering Science}, 62(13):3443--3454, 2007.

\bibitem{Microfluidics_Review}
T.~M. Squires and S.~R. Quake.
\newblock {Microfluidics: Fluid physics at the nanoliter scale}.
\newblock {\em Rev. Mod. Phys.}, 77(3):977, 2005.

\bibitem{GasFlows_Nicolas}
N.~G. Hadjiconstantinou.
\newblock {The limits of Navier-Stokes theory and kinetic extensions for
  describing small-scale gaseous hydrodynamics}.
\newblock {\em Physics of Fluids}, 18(11):111301, 2006.

\bibitem{FluctuatingHydro_Garcia}
J.~B. Bell, A.~Garcia, and S.~A. Williams.
\newblock {Numerical Methods for the Stochastic Landau-Lifshitz Navier-Stokes
  Equations}.
\newblock {\em Phys. Rev. E}, 76:016708, 2007.

\bibitem{FluctuatingHydro_Coveney}
G.~De Fabritiis, M.~Serrano, R.~Delgado-Buscalioni, and P.~V. Coveney.
\newblock {Fluctuating hydrodynamic modeling of fluids at the nanoscale}.
\newblock {\em Phys. Rev. E}, 75(2):026307, 2007.

\bibitem{StagerredFluctHydro}
N.~K. Voulgarakis and J.-W. Chu.
\newblock {Bridging fluctuating hydrodynamics and molecular dynamics
  simulations of fluids}.
\newblock {\em J. Chem. Phys.}, 130(13):134111, 2009.

\bibitem{LLNS_S_k}
A.~Donev, E.~Vanden-Eijnden, A.~L. Garcia, and J.~B. Bell.
\newblock {On the Accuracy of Explicit Finite-Volume Schemes for Fluctuating
  Hydrodynamics}.
\newblock Preprint, {\tt arXiv:0906.2425}, 2009.

\bibitem{FluidMixing_DSMC}
K.~Kadau, C.~Rosenblatt, J.~L. Barber, T.~C. Germann, Z.~Huang, P.~Carles, and
  B.~J. Alder.
\newblock {The importance of fluctuations in fluid mixing}.
\newblock {\em PNAS}, 104(19):7741--7745, 2007.

\bibitem{LatticeBoltzmann_Polymers}
O.~B. Usta, A.~J.~C. Ladd, and J.~E. Butler.
\newblock {Lattice-Boltzmann simulations of the dynamics of polymer solutions
  in periodic and confined geometries}.
\newblock {\em J. Chem. Phys.}, 122(9):094902, 2005.

\bibitem{StochasticImmersedBoundary}
P.~J. Atzberger, P.~R. Kramer, and C.~S. Peskin.
\newblock {A stochastic immersed boundary method for fluid-structure dynamics
  at microscopic length scales}.
\newblock {\em J. Comp. Phys.}, 224:1255--1292, 2007.

\bibitem{PolymerShear_MD}
C.~Aust, M.~Kroger, and S.~Hess.
\newblock Structure and dynamics of dilute polymer solutions under shear flow
  via nonequilibrium molecular dynamics.
\newblock {\em Macromolecules}, 32(17):5660--5672, 1999.

\bibitem{DSMCReview_Garcia}
F.~J. Alexander and A.~L. Garcia.
\newblock {The Direct Simulation Monte Carlo Method}.
\newblock {\em Computers in Physics}, 11(6):588--593, 1997.

\bibitem{DPD_DNA}
F.~Xijunand~N. Phan-Thien, S.~Chen, X.~Wu, and T.~Y. Ng.
\newblock {Simulating flow of DNA suspension using dissipative particle
  dynamics}.
\newblock {\em Physics of Fluids}, 18(6):063102, 2006.

\bibitem{DSMC_MPCD_Gompper}
M.~Ripoll, K.~Mussawisade, R.~G. Winkler, and G.~Gompper.
\newblock {Low-Reynolds-number hydrodynamics of complex fluids by
  multi-particle-collision dynamics}.
\newblock {\em Europhys. Lett.}, 68(1):106, 2004.

\bibitem{DSMC_MPCD_MD_Kapral}
S.~H. Lee and R.~Kapral.
\newblock {Mesoscopic description of solvent effects on polymer dynamics}.
\newblock {\em J. Chem. Phys.}, 124(21):214901, 2006.

\bibitem{SHSD}
A.~Donev, A.~L. Garcia, and B.~J. Alder.
\newblock {A Thermodynamically-Consistent Non-Ideal Stochastic Hard-Sphere
  Fluid}.
\newblock Preprint, {\tt arXiv:0908.0510}, 2009.

\bibitem{AMAR_Burgers}
J.~B. Bell, J.~Foo, and A.~L. Garcia.
\newblock {Algorithm refinement for the stochastic Burgers equation}.
\newblock {\em J. Comput. Phys.}, 223(1):451--468, 2007.

\bibitem{AMAR_DSMC}
A.~L. Garcia, J.~B. Bell, W.~Y. Crutchfield, and B.~J. Alder.
\newblock {Adaptive Mesh and Algorithm Refinement using Direct Simulation Monte
  Carlo}.
\newblock {\em J. Comp. Phys.}, 154:134--155, 1999.

\bibitem{FluctuatingHydro_AMAR}
S.~A. Williams, J.~B. Bell, and A.~L. Garcia.
\newblock {Algorithm Refinement for Fluctuating Hydrodynamics}.
\newblock {\em SIAM Multiscale Modeling and Simulation}, 6:1256--1280, 2008.

\bibitem{BreakupNanojets}
M.~Moseler and U.~Landman.
\newblock Formation, stability, and breakup of nanojets.
\newblock {\em Science}, 289(5482):1165--1169, 2000.

\bibitem{Nanojets_Eggers}
J.~Eggers.
\newblock Dynamics of liquid nanojets.
\newblock {\em Phys. Rev. Lett.}, 89(8):084502, 2002.

\bibitem{BrownianMotors}
C.~Van den Broeck, P.~Meurs, and R.~Kawai.
\newblock From maxwell demon to brownian motor.
\newblock {\em New Journal of Physics}, 7:10, 2005.

\bibitem{RayleighBernard_Fluctuations}
M.~Wu, G.~Ahlers, and D.S. Cannell.
\newblock Thermally induced fluctuations below the onset of
  {Rayleigh}-{B\'enard} convection.
\newblock {\em Phys. Rev. Lett.}, 75(9):1743--1746, 1995.

\bibitem{Kolmogorov_Fluctuations_1}
I.~Bena, M.~Malek Mansour, and F.~Baras.
\newblock Hydrodynamic fluctuations in the {Kolmogorov} flow: Linear regime.
\newblock {\em Phys. Rev. E}, 59(5):5503--5510, 1999.

\bibitem{Kolmogorov_Fluctuations_2}
I.~Bena, F.~Baras, and M.~Malek Mansour.
\newblock Hydrodynamic fluctuations in the {Kolmogorov} flow: Nonlinear regime.
\newblock {\em Phys. Rev. E}, 62(5):6560--6570, 2000.

\bibitem{Detonation_Fluctuations}
A.~Lemarchand and B.~Nowakowski.
\newblock Fluctuation-induced and nonequilibrium-induced bifurcations in a
  thermochemical system.
\newblock {\em Molecular Simulation}, 30(11-12):773--780, 2004.

\bibitem{AMAR_DSMC_SAMRAI}
S.~Wijesinghe, R.~Hornung, A.~L. Garcia, and N.~Hadjiconstantinou.
\newblock {Three-dimensional Hybrid Continuum-Atomistic Simulations for
  Multiscale Hydrodynamics}.
\newblock {\em Journal of Fluids Engineering}, 126:768--777, 2004.

\bibitem{DSMC_AED}
A.~Donev, A.~L. Garcia, and B.~J. Alder.
\newblock {Stochastic Event-Driven Molecular Dynamics}.
\newblock {\em J. Comp. Phys.}, 227(4):2644--2665, 2008.

\bibitem{EventDriven_Alder}
B.~J. Alder and T.~E. Wainwright.
\newblock {Studies in molecular dynamics. I. General method}.
\newblock {\em J. Chem. Phys.}, 31:459--466, 1959.

\bibitem{LB_SoftMatter_Review}
B.~Duenweg and A.~J.~C. Ladd.
\newblock {Lattice Boltzmann simulations of soft matter systems}.
\newblock {\em ArXiv e-prints}, 803, 2008.

\bibitem{LLNS_Mori}
K.~T. Mashiyama and H.~Mori.
\newblock {Origin of the Landau-Lifshitz hydrodynamic fluctuations in
  nonequilibrium systems and a new method for reducing the Boltzmann equation}.
\newblock {\em J. Stat. Phys.}, 18(4):385--407, 1978.

\bibitem{LLNS_Espanol}
P.~Español.
\newblock {Stochastic differential equations for non-linear hydrodynamics}.
\newblock {\em Physica A}, 248(1-2):77--96, 1998.

\bibitem{Landau:Fluid}
L.D. Landau and E.M. Lifshitz.
\newblock {\em Fluid Mechanics}, volume~6 of {\em Course of Theoretical
  Physics}.
\newblock Pergamon, 1959.

\bibitem{BergerAMR89}
M.~J. Berger and P.~Colella.
\newblock Local adaptive mesh refinement for shock hydrodynamics.
\newblock {\em J. Comp. Phys.}, 82:64--84, 1989.

\bibitem{AMR_Hyperbolic3D}
J.~Bell, M.~Berger, J.~Saltzman, and M.~Welcome.
\newblock {Three-Dimensional Adaptive Mesh Refinement for Hyperbolic
  Conservation Laws}.
\newblock {\em SIAM Journal on Scientific Computing}, 15(1):127--138, 1994.

\bibitem{CouplingAnalysis_Weiqing}
W.~Ren.
\newblock {Analytical and numerical study of coupled atomistic-continuum
  methods for fluids}.
\newblock {\em J. Comp. Phys.}, 227(2):1353--1371, 2007.

\bibitem{HybridMethods_Review}
H.~S. Wijesinghe and N.~G. Hadjiconstantinou.
\newblock {Discussion of hybrid atomistic-continuum methods for multiscale
  hydrodynamics}.
\newblock {\em International Journal for Multiscale Computational Engineering},
  2(2):189--202, 2004.

\bibitem{HMM_Fluids}
W.~Ren and W.~E.
\newblock {Heterogeneous multiscale method for the modeling of complex fluids
  and micro-fluidics}.
\newblock {\em J. Comp. Phys.}, 204(1):1--26, 2005.

\bibitem{DSMCHybrid_Boyd}
T.~E. Schwartzentruber and I.~D. Boyd.
\newblock {A hybrid particle-continuum method applied to shock waves}.
\newblock {\em J. Comp. Phys.}, 215(2):402--416, 2006.

\bibitem{MD_NS_Robbins}
X.~B. Nie, S.~Y. Chen, W.~N. E, and M.~O. Robbins.
\newblock {A continuum and molecular dynamics hybrid method for micro-and
  nano-fluid flow}.
\newblock {\em J. Fluid Mech.}, 500:55--64, 2004.

\bibitem{FluctuatingHydroMD_Coveney}
G.~Giupponi, G.~De Fabritiis, and P.~V. Coveney.
\newblock {Hybrid method coupling fluctuating hydrodynamics and molecular
  dynamics for the simulation of macromolecules}.
\newblock {\em J. Chem. Phys.}, 126(15):154903, 2007.

\bibitem{FluctuatingHydroHybrid_MD}
R.~Delgado-Buscalioni and G.~De Fabritiis.
\newblock {Embedding molecular dynamics within fluctuating hydrodynamics in
  multiscale simulations of liquids}.
\newblock {\em Phys. Rev. E}, 76(3):036709, 2007.

\bibitem{MD_NSF_Robbins}
J.~Liu, S.~Y. Chen, X.~B. Nie, and M.~O. Robbins.
\newblock {A continuum--atomistic simulation of heat transfer in micro-and
  nano-flows}.
\newblock {\em J. Comp. Phys.}, 227(1):279--291, 2007.

\bibitem{ChapmanEnskogBook}
S.~Chapman and T.~G. Cowling.
\newblock {\em {The Mathematical Theory of Non-Uniform Gases}}.
\newblock Cambridge Univ. Press, 1970.

\bibitem{ChapmanEnskog_Generation}
A.~L. Garcia and B.~J. Alder.
\newblock {Generation of the Chapman-Enskog distribution}.
\newblock {\em J. Comp. Phys.}, 140(1):66--70, 1998.

\bibitem{ChapmanEnskogFluxes}
S.~Chou.
\newblock {Kinetic Flux Vector Splitting for the Navier Stokes Equations}.
\newblock {\em J. Comp. Phys.}, 130:217--230, 1997.

\bibitem{VACF_2Divergence}
C.~P. Lowe and D.~Frenkel.
\newblock {The super long-time decay of velocity fluctuations in a
  two-dimensional fluid}.
\newblock {\em Physica A}, 220(3-4), 1995.

\bibitem{DSMC_CellSizeError}
F.~Alexander, A.~L. Garcia, and B.~J. Alder.
\newblock {Cell Size Dependence of Transport Coefficients in Stochastic
  Particle Algorithms}.
\newblock {\em Phys. Fluids}, 10:1540--1542, 1998.
\newblock Erratum: Phys. Fluids, 12:731-731 (2000).

\bibitem{DSMC_TimeStepError2}
N.~G. Hadjiconstantinou.
\newblock {Analysis of discretization in the direct simulation Monte Carlo}.
\newblock {\em Physics of Fluids}, 12:2634, 2000.

\bibitem{UnbiasedEstimates_Garcia}
A.~L. Garcia.
\newblock {Estimating hydrodynamic quantities in the presence of microscopic
  fluctuations}.
\newblock {\em Communications in Applied Mathematics and Computational
  Science}, 1:53--78, 2006.

\bibitem{FluctHydroNonEq_Book}
J.~M.~O.~De Zarate and J.~V. Sengers.
\newblock {\em {Hydrodynamic fluctuations in fluids and fluid mixtures}}.
\newblock Elsevier Science Ltd, 2006.

\bibitem{VACF_Alder}
B.~J. Alder and T.~E. Wainwright.
\newblock Decay of the velocity autocorrelation function.
\newblock {\em Phys. Rev. A}, 1(1):18--21, 1970.

\bibitem{BrownianLB_VACF}
M.~W. Heemels, M.~H.~J. Hagen, and C.~P. Lowe.
\newblock {Simulating Solid Colloidal Particles Using the Lattice-Boltzmann
  Method}.
\newblock {\em J. Comp. Phys.}, 164:48--61, 2000.

\bibitem{MPCD_VACF}
T.~Ihle and D.~M. Kroll.
\newblock {Stochastic rotation dynamics. II. Transport coefficients, numerics,
  and long-time tails}.
\newblock {\em Phys. Rev. E}, 67(6):066706, 2003.

\bibitem{FluctuatingHydro_FluidOnly}
N.~Sharma and N.~A. Patankar.
\newblock {Direct numerical simulation of the Brownian motion of particles by
  using fluctuating hydrodynamic equations}.
\newblock {\em J. Comput. Phys.}, 201:466--486, 2004.

\bibitem{BrownianSRD_Review}
J.~T. Padding and A.~A. Louis.
\newblock {Hydrodynamic interactions and Brownian forces in colloidal
  suspensions: Coarse-graining over time and length scales}.
\newblock {\em Phys. Rev. E}, 74(3):031402, 2006.

\bibitem{BrownianParticle_SIBM}
P.~J. Atzberger.
\newblock {Velocity correlations of a thermally fluctuating Brownian particle:
  A novel model of the hydrodynamic coupling}.
\newblock {\em Physics Letters A}, 351(4-5):225--230, 2006.

\bibitem{BrownianParticle_IncompressibleSmoothed}
T.~Iwashita, Y.~Nakayama, and R.~Yamamoto.
\newblock {A Numerical Model for Brownian Particles Fluctuating in
  Incompressible Fluids}.
\newblock {\em Journal of the Physical Society of Japan}, 77(7):074007, 2008.

\bibitem{SHSD_PRL}
A.~Donev, A.~L. Garcia, and B.~J. Alder.
\newblock {Stochastic Hard-Sphere Dynamics for Hydrodynamics of Non-Ideal
  Fluids}.
\newblock {\em Phys. Rev. Lett}, 101:075902, 2008.

\bibitem{FluctuationDissipation_Kubo}
R.~Kubo.
\newblock {The fluctuation-dissipation theorem}.
\newblock {\em Reports on Progress in Physics}, 29(1):255--284, 1966.

\bibitem{VACF_FluctHydro}
E.~H. Hauge and A.~Martin-Lof.
\newblock {Fluctuating hydrodynamics and Brownian motion}.
\newblock {\em J. Stat. Phys.}, 7(3):259--281, 1973.

\bibitem{BrownianCompressibility_Zwanzig}
R.~Zwanzig and M.~Bixon.
\newblock {Compressibility effects in the hydrodynamic theory of Brownian
  motion}.
\newblock {\em J. Fluid Mech.}, 69:21--25, 1975.

\bibitem{VACF_LagrangeTheory}
T.~V. Lokotosh and N.~P. Malomuzh.
\newblock {Lagrange theory of thermal hydrodynamic fluctuations and collective
  diffusion in liquids}.
\newblock {\em Physica A}, 286(3-4):474--488, 2000.

\bibitem{VACF_IncompressibleSmoothed}
T.~Iwashita, Y.~Nakayama, and R.~Yamamoto.
\newblock {Velocity Autocorrelation Function of Fluctuating Particles in
  Incompressible Fluids: Toward Direct Numerical Simulation of Particle
  Dispersions}.
\newblock {\em Progress of Theoretical Physics Supplement}, 178:86--91, 2009.

\bibitem{AdiabaticPiston_Lieb}
E.~Lieb.
\newblock {Some problems in statistical mechanics that I would like to see
  solved}.
\newblock {\em Physica A}, 263(1-4):491--499, 1999.

\bibitem{AdiabaticPiston_EDMD}
E.~Kestemont, C.~Van den Broeck, and M.~Malek Mansour.
\newblock The 'adiabatic' piston: And yet it moves.
\newblock {\em Europhysics Letters (EPL)}, 49(2):143--149, 2000.

\bibitem{AdiabaticPiston_VACF}
J.~A. White, F.~L. Roman, A.~Gonzalez, and S.~Velasco.
\newblock The \&ldquo;adiabatic\&rdquo; piston at equilibrium: Spectral
  analysis and time-correlation function.
\newblock {\em EPL (Europhysics Letters)}, 59(4):479--485, 2002.

\bibitem{Piston_DetHydro}
M.~Malek Mansour, A.~L. Garcia, and F.~Baras.
\newblock Hydrodynamic description of the adiabatic piston.
\newblock {\em Phys. Rev. E}, 73(1):016121, 2006.

\bibitem{Piston_RelaxationModel}
M.~Cencini, L.~Palatella, S.~Pigolotti, and A.~Vulpiani.
\newblock Macroscopic equations for the adiabatic piston.
\newblock {\em Phys. Rev. E}, 76(5):051103, 2007.

\bibitem{NonEqThermo_micro}
C.~Bustamante, J.~Liphardt, and F.~Ritort.
\newblock The nonequilibrium thermodynamics of small systems.
\newblock {\em Physics Today}, 58(7):43--48, 2005.

\bibitem{AdiabaticPiston_Langevin}
C.~Van den Broeck, E.~Kestemont, and M.~M. Mansour.
\newblock {Heat conductivity by a shared piston}.
\newblock {\em Europhysics Letters}, 56(6):771--777, 2001.

\bibitem{DSMC_Fluctuations_Shear}
A.~L. Garcia, M.~Malek Mansour, G.~C. Lie, M.~Mareschal, and E.~Clementi.
\newblock {Hydrodynamic fluctuations in a dilute gas under shear}.
\newblock {\em Phys. Rev. A}, 36(9):4348--4355, 1987.

\bibitem{DynamicStructureFactor_walls}
M.~M. Mansour, A.~L. Garcia, J.~W. Turner, and M.~Mareschal.
\newblock {On the scattering function of simple fluids in finite systems}.
\newblock {\em J. Stat. Phys.}, 52(1):295--309, 1988.

\bibitem{AMR_ReactionDiffusion_Atzberger}
P.~J. Atzberger.
\newblock {Spatially Adaptive Stochastic Numerical Methods for Intrinsic
  Fluctuations in Reaction-Diffusion Systems}.
\newblock Preprint.

\end{thebibliography}

\end{document}